\documentclass[aps,physrev,reprint,superscriptaddress,longbibliography,floatfix]{revtex4-2}

\usepackage{amsmath,amssymb,bm,graphicx}
\usepackage{hyperref}
\hypersetup{hidelinks}
\usepackage{xcolor}
\usepackage{placeins}
\usepackage{orcidlink}
\newcommand{\ket}[1]{\lvert #1\rangle}
\newcommand{\bra}[1]{\langle #1\rvert}
\newcommand{\Tr}{\operatorname{Tr}}
\newcommand{\Prob}{\operatorname{Pr}}
\newcommand{\ad}{\mathrm{ad}}
\newcommand{\sud}{\mathrm{sud}}
\newcommand{\fric}{\mathrm{fric}}
\newcommand{\cyc}{\mathrm{cyc}}

\begin{document}

\title{Probing quasiperiodic spectral accessibility with a finite-rate quantum Otto cycle}

\author{Shujie Cheng~\orcidlink{0009-0008-7953-5265}}
\affiliation{Xingzhi College, Zhejiang Normal University, Lanxi 321100, China}
\affiliation{Department of Physics, Zhejiang Normal University, Jinhua 321004, China}

\author{Lucas C. C\'eleri~\orcidlink{0000-0001-5120-8176}}
\email{lucas@qpequi.com}
\affiliation{Instituto de F\'isica, Universidade Federal de Goi\'as, 74690-900, Goi\^ania-GO, Brazil}
\affiliation{Instituto de Física de São Carlos, Universidade de São Paulo, CP 369, 13560-970, São Carlos, SP, Brasil}
\author{Norton G. de Almeida~\orcidlink{0000-0001-8517-6774}}

\affiliation{Instituto de F\'isica, Universidade Federal de Goi\'as, 74690-900, Goi\^ania--GO, Brazil}
\author{Gao Xianlong~\orcidlink{0000-0001-6914-3163}}
\email{gaoxl@zjnu.edu.cn}
\affiliation{Department of Physics, Zhejiang Normal University, Jinhua 321004, China}

\author{G. D. de Moraes Neto\,\orcidlink{0000-0003-4273-8380}}
\email{gdmneto@gmail.com}
\affiliation{Department of Fundamental Sciences, Hainan Bielefeld University of Applied Sciences, Danzhou, Hainan 578101, China}

\begin{abstract}
Quasiperiodic systems combine sharply different spectral and localisation regimes with a dense hierarchy of dynamical time scales. We show that a finite-rate quantum Otto cycle can use this structure as a thermodynamic probe, revealing which parts of the spectrum become dynamically accessible and thermodynamically relevant during a driven transformation. For a generalised Aubry-Andr\'e-Harper chain, the resulting response is not determined by the static quasiperiodic regime alone: extended, near-critical, and crossing protocols do not exhibit universal ordering of work fluctuations. Instead, the statistics reflect transition pathways that are simultaneously dynamically coupled, thermally occupied, and energetically significant. The same spectral accessibility separates thermodynamic recovery from microscopic adiabatic recovery. Positive mean work can reappear while the population transfer dynamics remain strongly nonadiabatic because finite-rate population transfer is redistributed toward energetically less costly channels rather than suppressed. The full work distribution reveals a further separation, with a finite rate window in which positive-work cycles form a majority although the mean work remains negative. This behaviour persists in finite-size analysis. A thermally weighted spectral susceptibility connects these effects to a distributed hierarchy of dynamically active spectral regions rather than to a single minimum-gap bottleneck. Finite thermal contacts preserve this picture while introducing a second time scale and intercycle memory. Quantum Otto cycles can therefore provide a fluctuation-resolved thermodynamic probe of quasiperiodic spectral accessibility, complementary to direct spectroscopic and eigenstate-resolved diagnostics.
\end{abstract}
\maketitle

\section{Introduction}

Quasiperiodic lattices occupy an intermediate position between periodic order and random disorder. They are deterministic and aperiodic, yet can support localisation, critical states, mobility edges, and fractal spectral structure without quenched randomness. Aubry-Andr\'e-Harper (AAH) models provide a paradigmatic one-dimensional setting for this physics, with generalised forms hosting extended, localised and critical regimes~\cite{AubryAndre1980,Biddle2011,Ganeshan2015,Liu2015,Tong2021}. Their non-uniform spectra and wave functions have motivated studies of localisation transitions, multifractality, mobility edges, and transport. The same structure also renders parameter driving non-trivial, as variations in the control parameters can reorganise both the spatial character of the eigenstates and the dynamical pathways connecting them.

This distinction becomes especially important away from equilibrium. Finite-rate ramps through localisation or critical regions can generate excitations when the evolving state cannot longer follow the instantaneous spectrum, whereas sudden quenches expose complementary information through return amplitudes and Loschmidt echoes~\cite{Dziarmaga2010,Polkovnikov2011,Sinha2019,Tong2021,Yang2017,Heyl2013}. These limits connect quasiperiodic dynamics to Kibble-Zurek-type excitation production during slow critical ramps and to return-probability diagnostics after quenches. More generally, they show that an equilibrium phase label alone does not determine the response to a time-dependent transformation, with the parameter path and the time available to traverse it being also part of the dynamics.

Quasiperiodic localisation is experimentally accessible in platforms that permit increasingly detailed control and readout. Matter-wave localisation has been observed in one-dimensional incommensurate optical lattices~\cite{Roati2008,Luschen2018,An2021}, and generalised AAH models have been realised in programmable superconducting-qubit arrays, where extended, critical, and localised dynamics can be distinguished through spatial propagation-based observables and participation-based observables~\cite{Li2023}. Such measurements provide microscopic information on the evolution of the state. They also motivate a complementary question: which signatures of this structured dynamics remain visible in coarser observables that do not require reconstructing spatial profiles, participation entropies, or individual eigenstates?

Quantum thermodynamic cycles provide one route to this question because unitary spectral dynamics is converted directly into energy exchange. Quasiperiodic working media have been proposed for steady-state heat engines that exploit mobility edges as energy-selective structures~\cite{Chiaracane2020} and for cyclic machines whose Hamiltonian is varied between different quasiperiodic regimes~\cite{XuCheng2025,Suo2026,Zhou2026}. The proposals of related localisation-based engines have connected the spectral structure to reliability and power~\cite{YungerHalpern2019}. These studies establish that localisation properties can reshape thermodynamic operating regimes. However, in cyclic settings, isolated work strokes have been treated mainly through limiting transformations in which the state is effectively frozen during a rapid parameter change or populations remain attached to instantaneous energy levels during asymptotically slow driving~\cite{Suo2026,Zhou2026}. The finite-rate regime between these limits contains additional information since the operating boundary can shift with stroke duration, whereas different parts of a structured spectrum contribute unequally to the resulting energy exchange.

Finite-time Otto dynamics is well developed for other working media, including explicit unitary strokes and incomplete thermalization~\cite{Rezek2006,Lee2020,Jiao2021,McKeever2025}, as well as critical driving and power optimization~\cite{Revathy2020,Arezzo2024,XuNeto2024RabiStark}. Fluctuation-resolved thermodynamics adds information that is absent from mean work alone. Analytically tractable two-level Otto engines make this structure explicit in closed work distributions~\cite{Jiao2021}, while nuclear magnetic resonance experiments have reconstructed joint work-heat statistics for a driven spin-$1/2$ Otto cycle~\cite{Denzler2024QST}. Finite-time engines can exhibit broad and asymmetric work distributions~\cite{Denzler2020,Anka2024}, threshold probabilities distinguish likely from average work extraction~\cite{Cavina2016}, and mean-to-fluctuation ratios provide separate measures of output precision~\cite{Miller2020,Shastri2022,Jaseem2023,DamasAQT2026}. More broadly, power, efficiency, and precision obey fluctuation tradeoffs~\cite{Pietzonka2018,MillerTUR2021}, while full counting statistics extend the description to long-time energy currents and cumulants~\cite{Esposito2009,MohantaAgarwalla2023}. Finite-rate work statistics have also been studied in quantum-critical ramps~\cite{Ma2025}. Particularly relevant here, the entropy of the work distribution in an AAH system can reveal localisation information that is not apparent in its lower moments~\cite{Kiely2023}. However, to our knowledge, cycle-resolved work fluctuations have not previously been analysed in a quasiperiodic quantum thermal machine. This raises a distinct question whether extended, critical, and localised regimes leave a simple ordering in thermodynamic fluctuations once finite-rate dynamics and thermal occupation act together.

Here, we treat the driving rate as a physical control parameter of a single-particle quantum Otto cycle. During each isolated work stroke, the generalised AAH Hamiltonian is continuously varied and the state is propagated through the evolving spectrum. The conventional thermodynamic mode diagram thereby becomes rate dependent, and a path that produces positive work in the slow-driving reference can become work-consuming at faster driving, and the corresponding operating boundary can move or disappear as the available stroke time changes. We refer to this rate-dependent classification as a \emph{finite-time operating diagram}. The finite-rate stroke is thus used as a probe of the transformation itself, connecting microscopic population transfer to mean thermodynamic output and to the complete work distribution.

We used \emph{spectral accessibility} operationally to describe the portion of the evolving spectrum that contributes appreciably to this finite-rate thermodynamic response. Accessibility is determined jointly by the dynamical coupling between instantaneous branches, their thermal occupation, and the energetic consequences of the resulting transitions. The Hamiltonian path determines which spectral structures are encountered, while the bath temperatures select which of them carry appreciable thermodynamic weight. The resulting probe is coarser than direct spectroscopy or eigenstate-resolved localisation diagnostics, but it retains information about which parts of the structured spectrum actually participate in the driven energy exchange.

\noindent\textbf{Contributions and scope.} Our results first show that the static quasiperiodic regime does not impose a universal ordering on finite-rate work fluctuations. Extended, near-critical, and crossing protocols exchange their fluctuation ordering between the slow-driving, finite-rate, and state-frozen regimes, while matched controls preserve the enhanced finite-rate response of the crossing path. The relevant distinction is therefore dynamical, with the statistics depending on which parts of the spectrum become accessible during the stroke, rather than on the static phase label alone. Along the crossing path, this produces two further separations. Positive mean work recovers, while the population-transfer map remains strongly nonadiabatic, with a transition-resolved decomposition showing that the transferred population shifts toward energetically less costly channels. In addition, over a finite interval of driving rates, positive-work cycles form a majority even though the mean work remains negative. This stochastic separation persists across the Fibonacci sizes $L=233,377,610,987$, while large branch displacements are increasingly associated with costly negative-work trajectories. At the microscopic level, a thermally weighted spectral susceptibility identifies a distributed hierarchy of dynamically active spectral regions that tracks excess-energy production. The narrowest isolated avoided crossing contributes only a small fraction of this response, so the operating boundary reflects the collective action of many thermally active transitions rather than a single minimum-gap bottleneck.

Finite thermal contacts extend this picture beyond the complete-reset cycle. Incomplete renewal introduces a second dynamical time scale, modifies work, power, efficiency, and fluctuations, and generates correlations between successive cycles. Repeated operation must therefore distinguish single-cycle work statistics from long-time current fluctuations ~\cite{Watanabe2017,Esposito2009}. Temperature and Hamiltonian controls provide complementary limits, with cooling narrowing the thermally active spectral sector and exposing slow critical dynamics, while sufficiently rapid strokes approach the state-frozen regime. The analysis concerns a single-particle generalised AAH chain, positive-temperature reservoirs, and two-projective-measurement work statistics. Principal rate scans and finite-size analysis use linear retraced work strokes. The finite-size results establish persistence over the range studied and are not intended as a thermodynamic-limit claim.

\section{Model, finite-time cycle, and thermodynamic metrics}
\label{sec:framework}

A finite-time Otto cycle contains two physically distinct clocks. During the isolated work strokes the Hamiltonian changes, and the relevant time is the duration available for the state to reorganise through the evolving spectrum. During thermal contacts, the Hamiltonian is fixed while the reservoirs redistribute populations, and the relevant time is the duration available for relaxation. To separate these clocks, we first take both contacts to the complete-reset limit and isolate the finite duration of the work strokes. Then, the finite thermalization times are restored as an independent control in Sec.~\ref{sec:contacts}.

\subsection{Quasiperiodic working medium and control paths}

The working medium is a single particle on an open quasiperiodic chain of $L$ sites. Both the hopping amplitudes and the onsite energies carry the same incommensurate modulation. Changing the external controls does more than shift an overall energy scale, it reshapes the instantaneous spectrum and the spatial character of its eigenstates while the particle is being driven. We use the diagonal-plus-off-diagonal Aubry-Andr\'e-Harper Hamiltonian
\begin{align}
H(r,\theta)={}&-\sum_{n=1}^{L-1}J_n(r,\theta) \bigl(\ket{n}\bra{n+1}+\mathrm{H.c.}\bigr)\nonumber\\
&+V(r,\theta)\sum_{n=1}^{L}\cos(2\pi\alpha n+\phi)\ket{n}\bra{n},
\label{eq:hamiltonian}
\end{align}
where the bond-dependent hopping is \(J_n(r,\theta)=J+\delta J(r,\theta)\cos(2\pi\alpha n+\phi)\). We parametrize the modulation amplitudes as \(\delta J=r/\sqrt{1+\tan^2\theta/4}\) and \(V=\delta J\tan\theta\), so that \(r^2=\delta J^2+(V/2)^2\) and \(V/\delta J=\tan\theta\). The radial coordinate $r$ sets the overall modulation strength, while $\theta$ determines how that modulation is shared between the hop and onsite terms. The phase $\phi$ changes the quasiperiodic pattern relative to the open boundaries and $\alpha$ fixes its incommensurate spatial frequency. Unless otherwise stated, we use $J=\hbar=k_{\mathrm B}=1$, $\alpha=(\sqrt5-1)/2$, and $\phi=0$. The quantity $L$ counts the lattice sites and is also the dimension of the Hilbert space of one-particle.

The static model contains extended, critical, and localised regimes~\cite{Liu2015,Suo2026,Lu2025,Ganeshan2015}, summarised schematically in Fig.~\ref{fig:workingmedium_cycle}(a) for orientation. The finite-time calculation compares three paths within this same Hamiltonian family, where one remains on the extended side, another approaches the static boundary, and the third crosses it. These paths were chosen to probe progressively more structured spectral evolution. The actual hierarchy of narrow avoided crossings encountered by the crossing path is established in Sec.~\ref{sec:spectral} and Appendix~\ref{app:accessibility_support}, rather than assumed from the endpoint labels. Their precise endpoints are shown in Fig.~\ref{fig:workingmedium_cycle}(a). Unless otherwise stated, we fix $\theta=\pi/4$ and $T_c/J=0.1$. The bath temperature and the rate of work-stroke are treated as independent control parameters. Because these representative protocols have different excursion lengths, comparisons intended to isolate the role of the static spectral regime are performed separately using equal-excursion controls. 

\subsection{Finite-time Otto cycle and thermodynamic metrics}

\begin{figure*}[t]
\centering
\includegraphics[width=\textwidth]{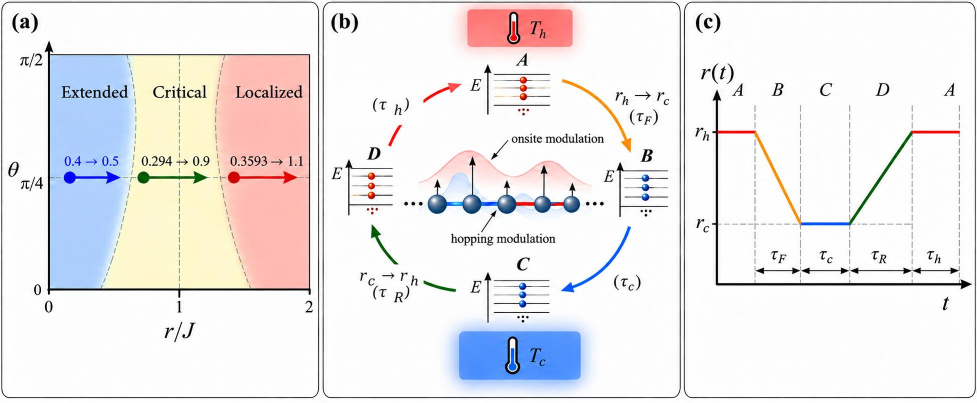}
\caption{\textbf{Working medium, static structure, and finite-time Otto protocol.} (a) Schematic summary of the static phase structure of the diagonal-plus-off-diagonal quasiperiodic model following Refs.~\cite{Liu2015,Suo2026}. The blue, pale-yellow, and red shaded regions denote the extended, critical, and localised regimes, respectively. The arrows at $\theta=\pi/4$ mark the representative paths $0.4\to0.5$, $0.294\to0.9$, and $0.3593\to1.1$. (b) The single-particle quasiperiodic chain is embedded in the four-corner Otto cycle. The central cartoon distinguishes onsite modulation from bond-dependent hopping. The work strokes $A\to B$ and $C \to D$ change $r$, while $B \to C$ and $D \to A$ are the cold and hot thermal contacts. (c) Control profile $r(t)$ over one cycle. The work strokes have durations $\tau_F$ and $\tau_R$, while the cold and hot contacts last $\tau_c$ and $\tau_h$, respectively, in the finite-contact extension. Times are measured in units $\hbar/J$.}
\label{fig:workingmedium_cycle}
\end{figure*}

Figure~\ref{fig:workingmedium_cycle} embeds the working medium in the four-stroke Otto protocol. During $A \to B$ and $C \to D$ the particle is isolated from the baths while $r$ is changed, while the hot and cold contacts repopulate the endpoint spectra at a fixed Hamiltonian.

We first take the contacts to the complete-reset limit, so each work stroke starts from the appropriate Gibbs state. This removes incomplete thermalization and inter-stroke memory from the first part of the analysis and isolates the effect of work-stroke duration. If $E_n^b$ denotes the energy-ordered levels at the end point $b\in\{h,c\}$ and $\beta_b=(k_{\mathrm B}T_b)^{-1}$, the Gibbs populations are $p_{bn}=e^{-\beta_b E_n^b}/Z_b$, with $Z_b=\sum_n e^{-\beta_b E_n^b}$. Corner $A$ is the hot Gibbs state of $H(r_h)$ and corner $C$ the cold Gibbs state of $H(r_c)$, while corners $B$ and $D$ are generated by unitary work strokes.

We take $Q_b>0$ when heat enters the working medium and $W_{\mathrm{out}}>0$ when the particle performs net work. If $U_X$ is the mean energy in corner $X$,
\begin{align}
Q_h&=U_A-U_D,\qquad Q_c=U_C-U_B,\nonumber\\
W_{\mathrm{out}}&=(U_A-U_B)+(U_C-U_D)=Q_h+Q_c .
\label{eq:firstlaw}
\end{align}
The signs of these mean energy flows distinguish the conventional operating modes summarised in Table~\ref{tab:operating_modes}.
\begin{table}[t]
\caption{\textbf{Thermodynamic operating modes.} A plus sign denotes energy entering the indicated system for heat, and positive $W_{\mathrm{out}}$ denotes work delivered by the working medium.}
\label{tab:operating_modes}
\begin{ruledtabular}
\begin{tabular}{lccc}
Mode & $W_{\mathrm{out}}$ & $Q_h$ & $Q_c$ \\
\hline
Engine    & $+$ & $+$ & $-$ \\
Refrigerator & $-$ & $-$ & $+$ \\
Heater    & $-$ & $-$ & $-$ \\
Accelerator & $-$ & $+$ & $-$ \\
\end{tabular}
\end{ruledtabular}
\end{table}
Within the engine sector, the mean efficiency is \(\eta=W_{\mathrm{out}}/Q_h\).

If only the endpoints were specified, this classification would say nothing about the speed with which the spectrum traversed~\cite{XuCheng2025,Suo2026,Zhou2026}. Here, $B$ and $D$ are generated by explicit time evolution, so shortening a work stroke can change not only the amount of work, but also whether the same endpoints operate as an engine at all. We use the term \emph{finite-time operating diagram} for this rate-dependent classification.

\subsection{Work-stroke dynamics: fast and slow limits}

We measure energies in units of $J$ and times in units of $\hbar/J$. Thus, a dimensionless stroke duration $\tau$ corresponds to a physical duration $\tau_{\rm phys}=(\hbar/J)\tau$. Throughout the numerical calculations, we set $J=\hbar=1$. For linear ramps,
\begin{align}
r_F(t)&=r_h+\frac{r_c-r_h}{\tau_F}t,
\qquad 0\le t\le\tau_F,\nonumber\\
r_R(t)&=r_c+\frac{r_h-r_c}{\tau_R}t,
\qquad 0\le t\le\tau_R ,
\label{eq:ramps}
\end{align}
For the symmetric protocol, \(\tau_F=\tau_R=|r_c-r_h|/\nu\), with \(\nu=|dr/dt|\). In these units, the driving rate is expressed as \(\nu/J\). The larger values correspond to shorter work strokes. Since the three paths have different excursion lengths, equal values of \(\nu/J\) do not correspond to equal stroke durations. Equivalently, the plotted rate $\nu/J$ is the dimensionless derivative $|d(r_{\rm phys}/J)/d(Jt_{\rm phys}/\hbar)|$. Platform-specific coherence and thermalization times enter after choosing the energy scale and reservoir rates of a concrete realisation.

Thermodynamic isolation and energy-ordered slow driving are distinct notions~\cite{Rezek2006,Lee2020}. In a very fast stroke the Hamiltonian changes before the wave function can reorganise appreciably in the site basis, defining the state-frozen reference. At the opposite extreme, an infinitely slow traversal follows the energy-ordered instantaneous eigenstates through the avoided-crossing structure and defines the strict slow-driving, energy-ordered reference. The quasiperiodic problem makes the distinction consequential because useful thermodynamic output can be recovered before this microscopic slow-driving limit is reached.

Let $U_F$ be the forward propagator and $\ket{n_h}$ and $\ket{m_c}$ the energy eigenstates of the endpoint. The forward population map is \(P_F(m|n)=|\langle m_c|U_F|n_h\rangle|^2\). For the reverse stroke, $P_R(\ell|k)=|\langle \ell_h|U_R|k_c\rangle|^2$. Because the Hamiltonian is real and the reverse stroke retraces the same path at the same rate, $P_R=P_F^{\mathsf T}$. In energy-ordered labels, $P_{\ad}(m|n)=\delta_{mn}$ is the strict slow-driving map, whereas $P_{\sud}(m|n)=|\langle m_c|n_h\rangle|^2$ is the state-frozen map. The open tridiagonal Hamiltonians have nondegenerate spectra over the paths considered, so instantaneous branches are ordered by eigenvalue. Near avoided crossings, their exchange of microscopic character is diagnosed by overlaps rather than by relabelling the energy order.

To locate the actual dynamics between these two limits, we use
\begin{equation}
D_{\ad}=\frac{\|P_F-I\|_F}{\sqrt L},\qquad
D_{\sud}=\frac{\|P_F-P_{\sud}\|_F}{\sqrt L}.
\label{eq:distances}
\end{equation}
Thus, $D_{\ad}=0$ means energy-ordered population following, while $D_{\sud}=0$ means state-frozen population dynamics. These are distances between population maps, not measures of phase coherence.

The Gibbs weights are branch dependent. To identify departures from energy-ordered following on the branches carrying thermal weight, we also use
\begin{align}
\ell_F&=\sum_n p_{hn}[1-P_F(n|n)],\nonumber\\
\ell_R&=\sum_m p_{cm}[1-P_R(m|m)],
\label{eq:leakage}
\end{align}
with $\bar\ell=(\ell_F+\ell_R)/2$. The map distances characterise the complete transition matrix, whereas $\bar\ell$ emphasises the thermally occupied sector. Together with the thermodynamic observables, these quantities separate operational recovery from the microscopic population as follows.

The work-stroke propagator is evaluated with unitary midpoint Cayley-Crank-Nicolson evolution. The convergence with respect to the time step and the independent adaptive propagation Schr\"odinger are documented in the Appendix~\ref{app:propagation}.

\subsection{Single-cycle work and heat statistics}

Even with fixed controls and bath temperatures, individual cycles exchange different amounts of energy because the thermal contacts sample different energy branches, and the finite-rate work strokes connect those branches probabilistically. Mean work and heat describe only the centre of a distribution of cycle outcomes. We resolve these outcomes with a TPM construction on each work stroke, combined into a four-corner cycle trajectory~\cite{Talkner2007,Campisi2011}. The resulting distribution is an operational measurement-resolved quantity. Projective energy readout dephases the state, so monitored and unmonitored coherent engines can carry different fluctuation statistics~\cite{Son2021Monitoring}. For the complete-reset cycle used below, the stroke-start states are Gibbs diagonal endpoint energy readout preserves the mean energy, and the following reset removes residual coherence. The quoted complete-reset mean work and heat are invariant under retaining or discarding the TPM record. With coherence-retaining finite contacts, the monitoring prescription instead becomes part of the dynamical specification.

Upon starting a complete-reset trajectory at the corner $C$, a reverse-stroke transition $\ket{k_c}\to\ket{\ell_h}$ is followed by the hot reset $\ket{\ell_h}\to\ket{n_h}$ and then the forward-stroke transition $\ket{n_h}\to\ket{m_c}$. The four-index record $\gamma=(k,\ell,n,m)$ is sufficient for work and hot heat. To identify the physical heat exchanged during the subsequent cold contact of the same realisation, append the reset outcome $\ket{m_c}\to\ket{k'_c}$ and write $\widetilde\gamma=(k,\ell,n,m,k')$. Then
\begin{align}
W_\gamma&=E_k^c-E_\ell^h+E_n^h-E_m^c,\nonumber\\
Q_{h,\gamma}&=E_n^h-E_\ell^h,\nonumber\\
Q_{c,\widetilde\gamma}^{\rm phys}&=E_{k'}^c-E_m^c,\qquad
\Delta E_{\widetilde\gamma}=E_{k'}^c-E_k^c,
\label{eq:trajectory_energetics}
\end{align}
so the trajectory first law under our output-work convention is
\begin{equation}
W_\gamma=Q_{h,\gamma}+Q_{c,\widetilde\gamma}^{\rm phys}
-\Delta E_{\widetilde\gamma}.
\label{eq:trajectory_first_law}
\end{equation}
For a complete reset, $k$ and $k'$ are independent draws from the same cold Gibbs distribution. Hence $\langle\Delta E\rangle=0$. The mean balance therefore reduces to $W_{\rm out}=Q_h+Q_c$, while the work distribution and the joint work-hot-heat statistics require only the marginal four-index $\gamma$.

Complete resets make the hot and cold preparations statistically independent, so \(\Pi_\gamma=p_{ck}P_R(\ell|k)\,p_{hn}P_F(m|n)\). The resulting joint distribution is
\begin{align}
P_{\cyc}(W,Q_h)=\sum_{k,\ell,n,m}&\Pi_\gamma\,
\delta\!\left(W-W_\gamma\right)\nonumber\\
&\times\delta\!\left(Q_h-Q_{h,\gamma}\right),
\label{eq:joint_tpm}
\end{align}
and the work distribution is its marginal. Its first moments reproduce the thermodynamic averages $\langle W\rangle=W_{\mathrm{out}}$ and $\langle Q_h\rangle=Q_h$, so the mean efficiency of the engine remains $\eta=\langle W\rangle/\langle Q_h\rangle$. We retain the joint distribution when work-heat correlations are needed, but do not use the trajectory ratio $W_\gamma/Q_{h,\gamma}$ as a primary observable because trajectories with $Q_{h,\gamma}=0$ make that ratio singular~\cite{Denzler2020,Anka2024}.

For the finite-time engine, the most transparent work statistics are
\begin{align}
P_+&=\Prob(W>0),\qquad
\sigma_W=\sqrt{\left\langle (W-\langle W\rangle)^2\right\rangle},\nonumber\\
\mathcal R&=\frac{\langle W\rangle}{\sigma_W}.
\label{eq:workmetrics}
\end{align}
They answer different operational questions: $P_+$ is the probability that one cycle delivers positive work~\cite{Cavina2016}, $\sigma_W$ sets the fluctuation scale and $\mathcal R$ compares the mean output with that width. For $\langle W\rangle>0$, we refer to $\mathcal R$ as the reliability of the engine~\cite{Miller2020,Shastri2022,Jaseem2023,DamasAQT2026}. Otherwise, it is retained only as a signed fluctuation diagnostic. These observables place mean thermodynamics, microscopic population transfer, and trajectory-level fluctuations on the same finite-rate axis.

\begin{figure*}[t]
\centering
\includegraphics[width=\textwidth]{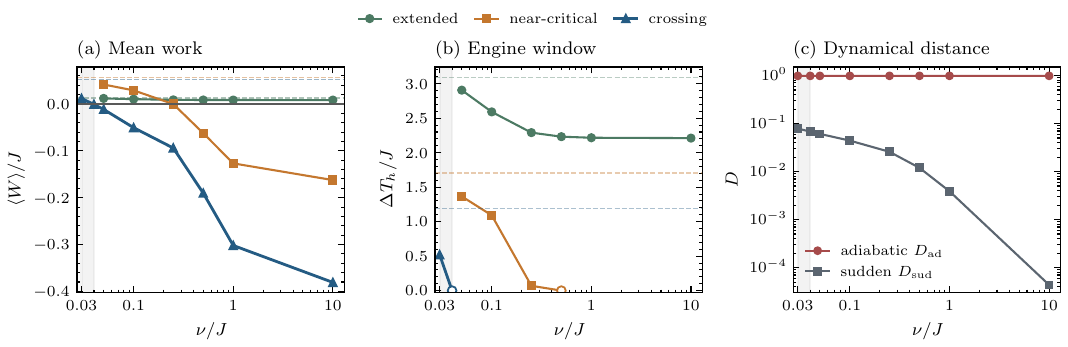}
\caption{\textbf{Finite-rate response of the representative paths.} (a) Mean output work versus driving rate. The extended, near-critical, and crossing curves use their representative hot-bath settings $T_h/J=1.0$, $0.8$, and $0.65$, respectively. (b) Width $\Delta T_h/J$ of the hot-bath temperature interval supporting engine operation. Faint dashed lines mark the strict energy-ordered widths. The open marker identifies the numerically resolved onset of window closure, with no engine window reappearing at higher evaluated rates. (c) Crossing-path distances from the strict energy-ordered and state-frozen population maps, together with the thermally weighted leakage $\bar\ell$, shown on a logarithmic scale. The shaded band marks the displayed $0.03\leq\nu/J\leq0.04$ interval bracketing recovery of positive mean work on the crossing path.}
\label{fig:access}
\end{figure*}

\section{Finite-rate accessibility and its spectral origin}
\label{sec:spectral}

How long must the isolated work stroke be before the engine operates again, and is that duration also sufficient for microscopic energy-ordered following? The crossing path makes these questions distinct because its thermally occupied states traverse a structured set of avoided crossings. Figure~\ref{fig:access} compares the thermodynamic and microscopic responses on the same driving-rate axis.

At an avoided crossing, an energy-ordered eigenstate can remain continuous, while its wave-function character is exchanged with a neighbouring branch. In the strict energy-ordered slow-driving limit, the state follows the evolving eigenstate through that exchange; a faster traversal can preserve more of the incoming microscopic character. For a single isolated crossing, this competition is captured by Landau--Zener physics~\cite{Zener1932}. However, along the quasiperiodic ramp, thermally occupied states encounter many crossings with widely different gaps and occupations.

The finite-temperature engine response samples a weighted collection of branches, each undergoing its own sequence of transfers. Appendix~\ref{app:accessibility_support} resolves representative branch exchange, lower-tail spacing statistics, and a thermally weighted spectral susceptibility. Away from exact degeneracies, this susceptibility is proportional to a Gibbs-population-weighted quantum metric. At $\nu/J=0.04$, the $20\%$ of forward-stroke intervals with the highest susceptibility carry a positive excess-energy growth $62.8\%$. An exact additive decomposition of both work strokes gives a broader cycle-level response: the top $20\%$ of the path ranked by the combined forward--reverse susceptibility carries $31.6\%$ of the total work deficit, while the window $|r-r_*|/J\leq0.005$ around the narrowest isolated avoided crossing contributes only $2.01\%$. The finite-rate response is therefore distributed across thermally occupied, dynamically coupled spectral regions spanning a broad hierarchy of gaps.

The susceptibility also guides a fixed-duration control. Keeping the total stroke time at $\tau=18.5167$ and reallocating $25\%$ more time to the highest-susceptibility $20\%$ of the path, compensated for by $25\%$ less time in the lowest-susceptibility $20\%$, changes the crossing-path work from $\langle W\rangle/J=-6.98\times10^{-4}$ for the uniform ramp to $+3.12\times10^{-3}$. Reversing the allocation gives $-4.79\times10^{-3}$. In all three cases $D_{\ad}\simeq0.983$, so the population map remains far from the energy-ordered following. The susceptibility therefore identifies where an increment of stroke time has the largest thermodynamic effect, while the detailed transition probabilities retain their dependence on dynamical phases and interference.

Figure~\ref{fig:access}(a) shows that shortening the work stroke has little effect on the extended-side path over most of the displayed range, while the near-critical and crossing paths lose output much earlier. The three curves use different representative hot-bath temperatures, so they compare finite-rate sensitivity along the chosen paths. In common $T_h/J=0.65$, $T_c/J=0.1$ and stroke duration $\tau=18.5167$, the extended and near-critical paths give $\langle W\rangle/J\simeq0.0130$ and $0.0465$, while the crossing path is slightly work-consuming, $\langle W\rangle/J\simeq-6.99\times10^{-4}$, and has the largest work width. The comparison of equal-excursions in Appendix~\ref{app:accessibility_support} gives the same qualitative distinction. These controls identify a path-dependent enhancement of the finite-rate response.

The width of the engine-window in Fig.~\ref{fig:access}(b) removes that single-temperature restriction. The extended path retains a broad hot-temperature interval, the near-critical window contracts, and the crossing-path window is the first to collapse. For the crossing path, increasing the duration of the stroke from $\tau\simeq18.5$ in $\nu/J=0.04$ to $\tau\simeq24.7$ in $0.03$ restores a finite engine window. These durations follow from $|r_c-r_h|/J\simeq0.7407$ and are quoted in the dimensionless time convention of Sec.~\ref{sec:framework}; the denser recovery brackets are listed in Appendix~\ref{app:accessibility_support}, Table~\ref{tab:app_certification}.

The operating-mode map is dominated by an engine-to-non-engine crossover. For the crossing protocol, the operation of the refrigerator is limited to a narrow low-temperature interval in the strict slow-driving reference, approximately $0.15\lesssim T_h/J\lesssim0.185$, and is absent in every finite driver evaluated. At $\nu/J=0.03$ the finite-rate diagram still contains an engine sector, while the remaining points are accelerator or heater operation; from $\nu/J=0.04$ onwards the evaluated crossing-path diagram contains only the latter non-engine modes. The finite-time boundary studied here is primarily an engine-to-non-engine crossover; refrigeration remains confined to the narrow low-temperature slow-driving sector.

The microscopic response evolves on a much slower scale. Figure~\ref{fig:access}(c) shows that $D_{\sud}$ changes by orders of magnitude as the stroke is lengthened, so the dynamics is progressively leaving the state-frozen limit, while $D_{\ad}$ remains of order unity throughout the displayed range. At $\nu/J=0.03$, where the engine operation has returned, $D_{\sud}\simeq0.079$ and $D_{\ad}\simeq0.982$. The Gibbs-weighted leakage is simultaneously $\bar\ell\simeq0.966$, confirming that the thermally occupied branches themselves remain strongly displaced from the energy-ordered following.

The distances from the population-map and leakage quantify redistribution, whereas the work deficit also weights the energies carried by those transfers. Resolving the transition $W_{\ad}-W_\nu$ by transition shows that, when the crossing protocol is slowed from $\nu/J=0.04$ to $0.03$, the work deficit decreases by approximately $24\%$, while $\bar\ell$ changes by less than $0.2\%$ and $D_{\ad}$ remains essentially unchanged. Most of the reduction comes from the reverse stroke and from changes among large branch-displacement transitions. Positive mean work recovers because the still strongly off-diagonal transferred population shifts toward channels with smaller work penalties. The exact decomposition and the rate-resolved values are given in the Appendix~\ref{app:transition_work}.

Engine operation is restored at work-stroke durations of order tens of $\hbar/J$, while the complete transition map and the Gibbs-weighted branch dynamics remain far from energy-ordered following. The work distribution provides an additional level of information, distinguishing positive mean output from typical-cycle behavior and fluctuation-based reliability.

\begin{figure*}[t]
\centering
\includegraphics[width=0.96\textwidth]{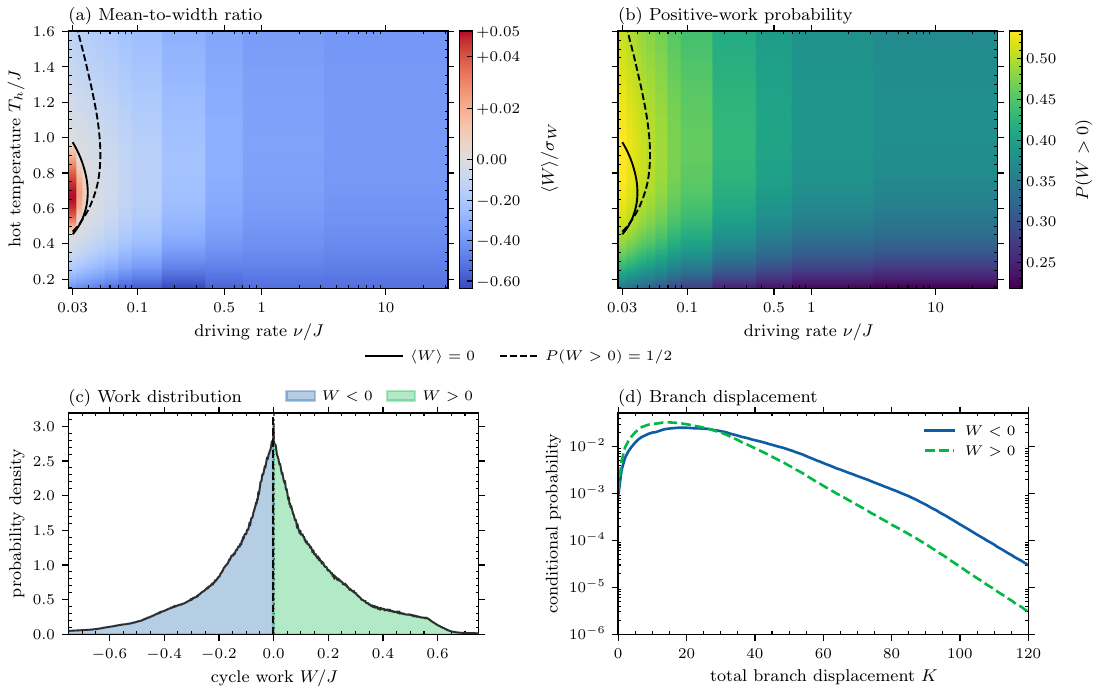}
\caption{\textbf{Stochastic operation and branch redistribution for the crossing path.}
	(a) Signed mean-to-width ratio $\langle W\rangle/\sigma_W$ over the $(T_h/J,\nu/J)$ plane.
	(b) Positive-work probability $P(W>0)$ over the same parameter range.
	In panels (a,b), the solid contour marks the mean-work boundary $\langle W\rangle=0$, while the dashed contour marks the majority-positive boundary $P(W>0)=1/2$; contours interpolate between the computed driving rates as visual guides.
	(c) TPM cycle-work distribution at $T_h/J=0.65$ and $\nu/J=0.04$, with negative- and positive-work sectors shaded separately. The dotted line marks $W=0$ and the dashed line the mean work.
	(d) Distribution of the total branch displacement $K$, conditioned on the sign of the cycle work, at the same operating point.}
\label{fig:work}
\end{figure*}

\section{Work fluctuations and distinct finite-rate criteria}
\label{sec:workstats}

\subsection{From mean operation to cycle-by-cycle reliability}

Positive mean work and typical-cycle output recover at distinct rates. The contours in the two heat maps of Fig.~\ref{fig:work} are separated. In Fig.~\ref{fig:work}(b), the contour $P(W>0)=1/2$ extends beyond the contour $\langle W\rangle=0$, so a majority of cycles can already produce positive work while the mean is still negative. Figure~\ref{fig:work}(a) adds a second separation: throughout the displayed positive-mean region, the signed mean-to-width ratio remains below about $0.05$. Majority-positive behavior appears before positive mean work, while the recovered engine region remains strongly fluctuation dominated.

The difference between the first two criteria follows from
\begin{equation}
\langle W\rangle=P_+\langle W\mid W>0\rangle
 +(1-P_+)\langle W\mid W\leq0\rangle .
\label{eq:conditionalmean}
\end{equation}
A small majority of positive-work cycles can coexist with a negative mean when the negative-work histories carry larger losses. The representative point in Fig.~\ref{fig:work}(c) makes this imbalance visible and locates it within the finite-rate quasiperiodic dynamics. At $T_h/J=0.65$ and $\nu/J=0.04$, just over half of the cycles have $W>0$, yet the mean remains slightly negative. Slowing the same path to $\nu/J=0.03$ restores a positive mean work, with engine reliability $\mathcal R\simeq0.049$; therefore, fluctuations remain large compared to the mean. The unbinned TPM construction used for the quoted moments, sign probabilities, and conditional means is given in Appendix~\ref{app:tpm}.

The fluctuation scale depends on the driving protocol as well as the static spectral regime. Equal-excursion controls at common bath temperatures separate this question from the different path lengths of the representative protocols. For $\Delta r/J=0.1$, the localised control has the largest $\sigma_W$ in the strict slow-driving reference, whereas the critical control is the widest in the state-frozen reference and at the representative finite rate $\nu/J=0.04$. A sliding equal-excursion scan places the strongest finite-rate response near the critical--localised crossover. The work-width ordering is protocol dependent: it changes between the slow-driving, finite-rate, and state-frozen protocols. The numerical controls are summarised in the Appendix Table~\ref{tab:equal_excursion_fluctuations}.
\subsection{Branch-redistribution structure of the negative-work tail}

To resolve the transition histories behind the negative-work tail, we define for a complete TPM cycle with branch changes $n\to m$ and $k\to\ell$ in the two work strokes the total displacement in the energy-ordered branch index as \(K=|m-n|+|\ell-k|\). This is a spectral-branch displacement, not a spatial localisation measure. Large $K$ means that the two work strokes carry the trajectory far from the branches it would follow in the energy-ordered limit.

Figure~\ref{fig:work}(d) shows that the negative-work sector has a significantly heavier large tail $K$. At the representative point, $13.28\%$ of negative-work trajectories have $K\geq50$, compared to $4.24\%$ of positive-work trajectories, and this same tail carries $27.98\%$ of the total negative-work loss. The energetic severity also increases with displacement: the mean loss is about $0.11J$ for $K<10$, $0.31J$ for $50\leq K<60$, $0.43J$ for $60\leq K<80$, and $0.83J$ for $K\geq100$. The large enrichment $K$ persists throughout the threshold range $20\leq K_0\leq100$, while the conditional energetic loss grows with displacement. Appendix Table~\ref{tab:tpm_tail_energetics} gives the threshold-resolved values.

The same redistribution signature also appears on smoother paths, but the crossing brings it into relevance at much slower driving. At $\nu/J=0.04$, the mean negative-work displacement is almost branch preserving on the extended path, $\langle K\mid W<0\rangle\simeq0.2$, but is already about $29$ on the crossing path. The near-critical control reaches a comparable redistribution only near $\nu/J=10$. For the crossing path, the negative tail of the large-displacement becomes prominent at a driving rate roughly $250$ times slower than for the near-critical control.

The negative-work sector is concentrated in the thermally occupied low-energy manifold. At $\nu/J=0.04$, the lowest 32 hot-side branches contain $50.8\%$ of the Gibbs population, but generate $71.1\%$ of the negative cycles and $61.4\%$ of the total loss of work. negative. The effect involves many levels, but is strongly weighted toward the occupied low-energy manifold. Together with the displacement-resolved loss data, this links strong branch redistribution in the thermally populated sector directly to the negative-work tail.

\FloatBarrier

\section{Finite-size persistence and quasiperiodic arithmetic}
\label{sec:size}

A finite quasiperiodic chain carries two kinds of size dependence at once. Increasing \(L\) enlarges the one-particle spectrum and resolves progressively finer pieces of the incommensurate spatial structure. We follow the Fibonacci sequence \(L=233,377,610,987\) at fixed inverse-golden modulation and crossing protocol to determine how the fluctuation-dominated, strongly nonadiabatic regime evolves as that structure is resolved more finely. This finite-size question is distinct from previously studied reliability scaling in many-body-localised and finite-size Otto engines~\cite{YungerHalpern2019,DamasFiniteSize2026}.

\subsection{Persistence with increasing system size}

Table~\ref{tab:size_persistence} compares the complete work statistics available at the two endpoint sizes with transition diagnostics at all four Fibonacci sizes. The endpoint stochastic comparison changes very little between \(L=233\) and \(987\): the work-distribution width is close to \(0.244J\), the probability of positive work stays near \(0.53\), and the signed mean-to-width ratio is about \(0.05\). The broad, fluctuation-dominated cycle statistics established in Sec.~\ref{sec:workstats} are essentially unchanged between the two endpoint sizes.

The microscopic response is equally persistent. Across the four sizes, the distance from the state-frozen map decreases while the distance from the energy-ordered map and the thermally weighted leakage remain close to unity. The larger systems retain strongly nonadiabatic population transfer on the reservoir-populated branches at fixed rate, with increasing distance from energy-ordered following and decreasing distance from the state-frozen map.

\begin{table}[b]
\caption{\textbf{Finite-size persistence at \(\nu/J=0.03\).} Work statistics are shown at the endpoint sizes \(L=233\) and \(987\), while transition diagnostics span the full Fibonacci sequence considered.}
\label{tab:size_persistence}
\squeezetable
\begin{ruledtabular}
\begin{tabular}{lcccc}
 & \(233\) & \(377\) & \(610\) & \(987\) \\
\hline
\(\langle W\rangle/J\) & 0.0118 & -- & -- & 0.0132 \\
\(\sigma_W/J\) & 0.2428 & -- & -- & 0.2438 \\
\(\mathcal R\) & 0.049 & -- & -- & 0.054 \\
\(P(W>0)\) & 0.5260 & -- & -- & 0.5279 \\
skewness & -0.554 & -- & -- & -0.546 \\
excess kurtosis & 3.275 & -- & -- & 3.263 \\
\hline
\(D_{\ad}\) & 0.9825 & 0.9876 & 0.9920 & 0.9952 \\
\(D_{\sud}\) & 0.0787 & 0.0619 & 0.0488 & 0.0384 \\
\(\bar\ell\) & 0.9659 & 0.9762 & 0.9854 & 0.9910 \\
\end{tabular}
\end{ruledtabular}
\end{table}

\subsection{Boundary shifts and the role of quasiperiodic arithmetic}

The precise boundaries shift with size, but the interval between them persists. At the common rate $\nu/J=0.041$, all four sizes satisfy $\langle W\rangle<0$ and $P(W>0)>1/2$: the mean work changes from $-1.96\times10^{-3}J$ at $L=233$ to $-4.91\times10^{-4}J$ at $L=987$, and $P_+$ changes from $0.5043$ to $0.5064$. Bracketing the two zero crossings separately shows that the estimated width of the sign-separated interval remains approximately $4.2$--$4.4\times10^{-3}$ in $\nu/J$ across $L=233,377,610,987$, even though both boundary locations move. The majority-positive and positive-mean criteria retain the same ordering and a finite separation across the four sizes despite the shifts in boundary location. Appendix Table~\ref{tab:size_sign_boundaries} gives the four-size brackets and the large-$L$ convergence result; Fig.~\ref{fig:app_numerical_certification}(b) shows that the microscopic nonadiabaticity persists across the same sequence.

For the inverse-golden modulation, the Fibonacci size sequence also has an arithmetic meaning. The chain length is increased while the irrational frequency \(\alpha=(\sqrt5-1)/2\) itself is held fixed, and the chosen \(L\) values follow the natural denominators associated with its continued-fraction structure. Independent rational-approximant calculations show that the TPM observables approach the inverse-golden result smoothly even as the smallest instantaneous spacing varies nonmonotonically.

Changing \(\alpha\) is a different physical operation from increasing \(L\). The silver-ratio control preserves the qualitative separation between majority-positive and positive-mean operations, whereas other irrational choices can modify the endpoint thermodynamics of the same nominal ramp. The inverse-golden Fibonacci sequence establishes finite-size persistence of the stochastic regime while retaining sensitivity to quasiperiodic frequency, Hamiltonian path, and thermal occupation as separate controls.

\section{Control parameters, temperature, and critical-dynamics limits}
\label{sec:controls}

The finite-rate mechanism relies on two distinct forms of spectral selection. Changing the quasiperiodic Hamiltonian changes the spectral structures traversed during the work strokes, whereas changing the bath temperatures changes the structures that carry appreciable thermodynamic weight. Figure~\ref{fig:controls_temperature} separates these two effects and connects them to the low-temperature critical-ramp and sudden-quench limits.

\subsection{Quasiperiodic and Hamiltonian controls}

The finite-rate stochastic separation persists across the phase offsets examined. generalised quasiperiodic models already show that transport and localisation diagnostics can vary strongly with modulation parameters~\cite{Lu2025}. Varying the quasiperiodic phase $\phi$ over the selected values produces only weak changes in the crossing-path response, and a silver-ratio control retains the qualitative separation between majority-positive and positive-mean operation. Changing the irrational frequency more broadly can alter the endpoint thermodynamics of the same nominal $r$ ramp and can even remove the strict energy-ordered engine. The response remains sensitive to the quasiperiodic frequency while varying only weakly across the phase realisations examined; the full scans are collected in Appendix~\ref{app:robustness}.

\begin{figure*}[t]
\centering
\includegraphics[width=0.98\textwidth]{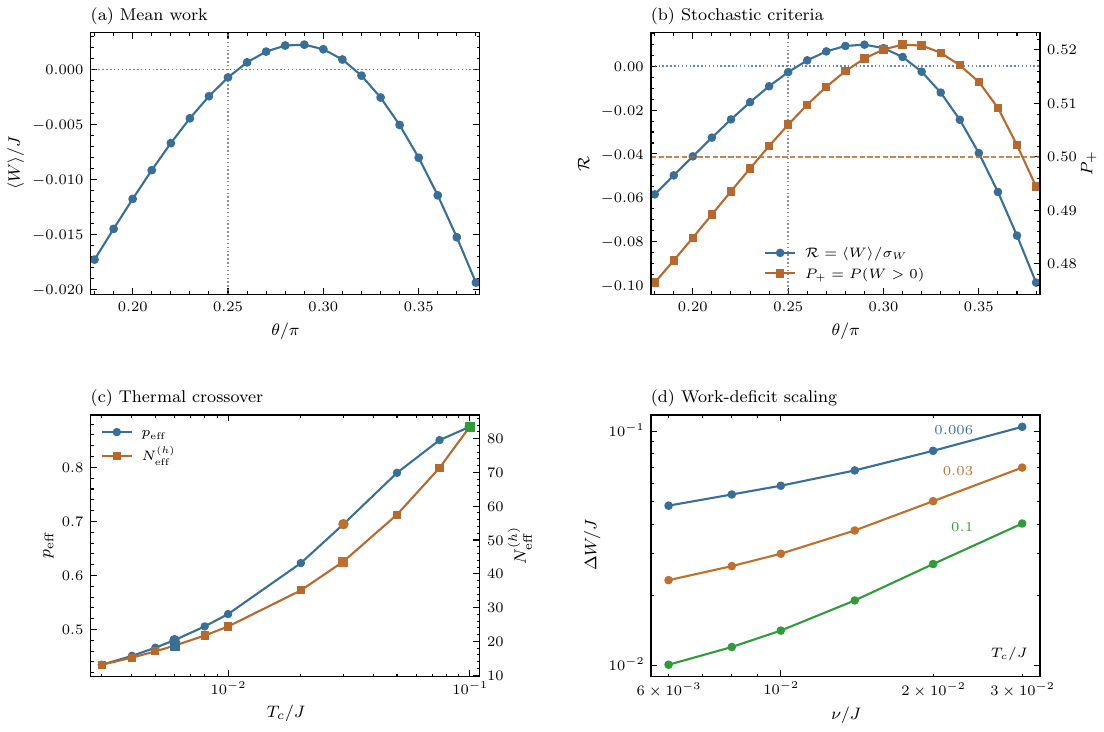}
\caption{\textbf{Hamiltonian-path control and the thermal crossover.}
(a) Mean work at $T_h/J=0.65$, $T_c/J=0.1$, and $\nu/J=0.04$ as the modulation angle $\theta$ changes the balance between diagonal and off-diagonal quasiperiodic modulation. The vertical dotted line marks the baseline path $\theta=\pi/4$.
(b) Signed mean-to-width ratio $\mathcal R=\langle W\rangle/\sigma_W$ and positive-work probability $P_+=P(W>0)$ for the same scan. The horizontal reference lines mark $\mathcal R=0$ and $P_+=1/2$.
(c) Effective finite-window work-deficit slope $p_{\rm eff}$ and hot-state participation $N_{\rm eff}^{(h)}=1/\sum_n p_{hn}^2$ when the two bath temperatures are lowered together at fixed ratio $T_h=6.5T_c$. The colored markers identify the temperatures used in panel (d).
(d) Raw work deficit $\Delta W=W_{\ad}-W_\nu$ versus rate at three representative temperatures, labeled directly by $T_c/J$. The curves show the slope change summarised in panel (c).}
\label{fig:controls_temperature}
\end{figure*}

The modulation angle $\theta$ is a more direct Hamiltonian control because it changes how the same radial amplitude is divided between diagonal and off-diagonal quasiperiodic modulation. Figure~\ref{fig:controls_temperature}(a) shows that this change alone can move the finite-rate crossing cycle into and out of positive mean work at $T_h/J=0.65$, $T_c/J=0.1$, and $\nu/J=0.04$. The operating interval is bounded on both sides: paths with too little or too much diagonal modulation become work-consuming even though the radial endpoints are unchanged.

The stochastic criteria sharpen this path dependence. In Fig.~\ref{fig:controls_temperature}(b), the signed mean-to-width ratio $\mathcal R$ follows the same broad optimum as the mean work, whereas $P_+$ continues to improve to a slightly different path before turning over. The displacement is small but systematic. The path that maximizes average output is slightly displaced from the path that maximizes the probability of positive work. At the same time, $\mathcal R$ stays small throughout the scan, so the entire comparison is fluctuation dominated.

\subsection{Temperature and the crossover in active spectral weight}

Temperature changes the thermodynamically active part of the spectrum without changing the Hamiltonian path. At the baseline temperatures $T_c/J=0.1$ and $T_h/J=0.65$, the hot Gibbs state occupies a broad manifold of branches, consistent with the many-level negative-work mechanism identified in Sec.~\ref{sec:workstats}. Lowering both reservoirs together at fixed ratio $T_h=6.5T_c$ suppresses the high-energy contribution and progressively restricts the cycle to the low-energy sector. Figure~\ref{fig:controls_temperature}(c) quantifies this contraction with the hot-state participation $N_{\rm eff}^{(h)}$, which falls from more than eighty effective branches at $T_c/J=0.1$ to fewer than twenty near $T_c/J=0.006$.

The finite-rate scaling changes at the same time. Over the fixed rate window used in Fig.~\ref{fig:controls_temperature}(d), the effective work-deficit slope evolves from $p_{\rm eff}\simeq0.88$ at $T_c/J=0.1$ toward $p_{\rm eff}\simeq0.48$ at $T_c/J=0.006$. The raw curves in panel (d) confirm that the change occurs in the slope of the work-deficit curve rather than only in isolated fit values. The concurrent decrease of $N_{\rm eff}^{(h)}$ and $p_{\rm eff}$ identifies a thermal crossover in which the thermally active spectral weight and the finite-rate response evolve together as the occupied energy window contracts. The participation ratio quantifies that contraction, while the effective exponent also depends on the detailed transition structure within the active window.

In the still lower-temperature limit, the thermodynamic interpretation becomes especially simple. The strict energy-ordered cycle returns the occupied ground branch to itself, whereas a finite-rate stroke creates excitation energies $\delta E_F$ and $\delta E_R$ on the forward and reverse ramps. The complete-reset cycle then obeys
\begin{equation}
W_{\ad}-W_\nu=\delta E_F+\delta E_R .
\label{eq:lowTwork}
\end{equation}
In this limit the work deficit equals the excitation energy produced by the two ramps, providing a direct bridge from a thermodynamic observable to critical-ramp excitation physics.

\subsection{Critical-ramp and sudden-quench limits}

The Kibble--Zurek mechanism provides a natural limiting language for slow driving across a continuous critical region: adiabatic following fails near the part of the ramp with the longest intrinsic response time, producing excitations with rate-dependent scaling~\cite{Dziarmaga2010,Polkovnikov2011,Sinha2019,Tong2021,Revathy2020}. For the quasiperiodic crossing path considered here, this picture is complicated by a hierarchy of narrow avoided crossings rather than a single isolated scale. A single-scale asymptotic interpretation would require an accessible slow-ramp window that is insensitive to this hierarchy; the present finite interval instead supports effective rate-dependent slopes.

Extending the joint cooling scan into the low-temperature regime yields a reproducible finite-window law. At $L=233$ and $T_c/J=0.006$, a fit over the full evaluated interval $0.006\leq\nu/J\leq0.030$ gives $p_{\rm eff}=0.480$. Nested four-point windows vary from $p_{\rm eff}=0.407$ on $0.006\leq\nu/J\leq0.014$ to $0.529$ on $0.010\leq\nu/J\leq0.030$, with intermediate windows lying between these values. Equation~\eqref{eq:lowTwork} confirms that the same rate dependence appears in the work deficit and in the total excitation energy. The systematic window drift, together with the size- and approximant-dependence of the quasiperiodic gap hierarchy, identifies these slopes as effective finite-window descriptors. All quoted low-temperature exponents refer to the $L=233$ scan. The resulting $p_{\rm eff}$ values provide a controlled finite-window connection to quasiperiodic Kibble--Zurek dynamics.

This distinction also separates the low-temperature critical-ramp limit from the finite-temperature engine boundary. The latter is reached while many thermally weighted branches remain active and before microscopic adiabatic following is restored. The former progressively isolates the lowest-energy excitation physics. Temperature connects these two regimes of the same driven Hamiltonian while changing the number and structure of the branches that carry thermodynamic weight.

The opposite limiting comparison is the sudden quench. In the state-frozen reference, the branch-resolved characteristic function develops the expected Loschmidt-type structure, and the crossing path shows a pronounced return minimum that is absent in smoother controls~\cite{Heyl2013,Heyl2018,Tong2021,Khan2024}. This provides a useful connection between the fast reference transformation and standard quench diagnostics.

Across the four sizes, the continuously driven characteristic-function minima saturate and the physical return probability remains finite. The finite-rate ramps occupy a nonsingular crossover regime, while the pronounced Loschmidt-type structure belongs to the sudden reference. The corresponding temperature, fit-window, and return diagnostics are given in Appendix~\ref{app:criticallimits}.

\section{Finite contacts, efficiency, power, and entropy production}
\label{sec:contacts}

The complete-reset cycle isolates the work-stroke dynamics by allowing each bath to restore its Gibbs populations fully. Finite thermal contacts restore the second clock introduced in Sec.~\ref{sec:framework}: the time required to renew those endpoint populations. Incomplete renewal leaves population memory in the stationary cycle, making thermalization itself a second accessibility constraint while also adding the contact duration to the performance problem~\cite{Jiao2021,DamasAQT2026,DamasFiniteSize2026}.

\subsection{Incomplete thermalization as a second accessibility constraint}

We model contact with bath $b\in\{h,c\}$ by the population semigroup
\begin{equation}
\bm p\longmapsto(1-\lambda_b)\bm p+\lambda_b\bm p_{\beta_b},
\qquad
\lambda_b=1-e^{-\Gamma_b\tau_b},
\label{eq:contact}
\end{equation}
where $\Gamma_b$ is the relaxation rate and $\tau_b$ the contact duration. The complete-reset reference is recovered only asymptotically as $\lambda_b\to1$. Equation~\eqref{eq:contact} is a single-rate Gibbs-renewal model; the numerical contact thresholds below therefore refer to this channel. A coherence-retaining Gibbs channel provides a complementary comparison of weak-contact sensitivity. The stationary population map and fixed-point construction are given in Appendix~\ref{app:contacts}.

Finite contacts introduce a second boundary in addition to the work-stroke boundary of Sec.~\ref{sec:spectral}. The unitary strokes may already be slow enough to support positive mean work, yet the stationary cycle can still fail if the baths do not renew the endpoint populations strongly enough. Table~\ref{tab:contact_thresholds} shows that the required symmetric reset strength depends strongly on the chosen finite-rate protocol. The near-critical control requires almost complete population renewal, whereas the extended and crossing protocols tolerate appreciably incomplete contacts. If either bath is fully reset, the required strength of the other is lower, showing that the two contacts jointly set the population memory retained by the stationary cycle.

\begin{table}[t]
\caption{\textbf{Finite-contact accessibility under the population semigroup.} The symmetric reset strength $\lambda_{\rm crit}$ marks the onset of positive stationary mean work for each representative finite-rate protocol.}
\label{tab:contact_thresholds}
\begin{ruledtabular}
\begin{tabular}{lccc}
Path & $\nu/J$ & $T_h/J$ & $\lambda_{\rm crit}$ \\
\hline
Extended & 0.25 & 1.0 & 0.46678 \\
Near-critical & 0.25 & 0.8 & 0.99995 \\
Crossing & 0.03 & 0.65 & 0.83861 \\
\end{tabular}
\end{ruledtabular}
\end{table}

Residual coherence can modify this contact boundary. To estimate its influence, we also propagate the full density matrix through the affine Gibbs channel $\rho\mapsto(1-\lambda)\rho+\lambda\rho_\beta$, which retains a fraction $1-\lambda$ of the pre-contact coherence. On the extended path at symmetric $\lambda=0.40$, the population model gives $W/J=-7.02\times10^{-4}$ whereas the coherence-retaining channel gives $W/J=1.89\times10^{-3}$, reversing the operating sign. On the crossing path both models change sign within the same numerical bracket $0.83<\lambda_{\rm crit}<0.84$; the population value in Table~\ref{tab:contact_thresholds} is $0.83861$. Thus the crossing-path contact boundary is insensitive to this particular coherence control at the quoted resolution, while the smoother path is qualitatively coherence sensitive in the weak-contact regime. Equation~\eqref{eq:contact} remains the population-level model used for the quantitative operating analysis. Appendix~\ref{app:contacts} gives the coherence-retaining channel and its convergence details.

\subsection{Power, stochastic accessibility, and irreversibility}

A finite contact must be paid for in cycle time. Defining the contact exposure $x_b=\Gamma_b\tau_b$, the stationary cycle duration is
\begin{equation}
\tau_{\cyc}=\frac{2|r_c-r_h|}{\nu}
+\frac{x_h}{\Gamma_h}+\frac{x_c}{\Gamma_c},
\label{eq:cycletime}
\end{equation}
and the mean power is $\mathcal P=W_{\rm out}/\tau_{\cyc}$ in the engine sector. With time measured in units of $\hbar/J$, the cycle duration in Eq.~\eqref{eq:cycletime} converts directly to physical time once $J$ and the reservoir relaxation rates are specified. The plotted bath rate $\Gamma_b/J$ corresponds to the dimensionless physical rate $\hbar\Gamma_{b,\rm phys}/J$. For example, the crossing path has $\tau\simeq24.7$ at $\nu/J=0.03$ and $\tau\simeq43.6$ at the $L=233$ grid maximum-power rate $\nu/J=0.017$. For $\hbar\Gamma_{b,\rm phys}/J=1$, a contact exposure $x_b=5$ corresponds to a contact duration $5\hbar/J$.

\begin{figure*}[t]
\centering
\includegraphics[width=0.98\textwidth]{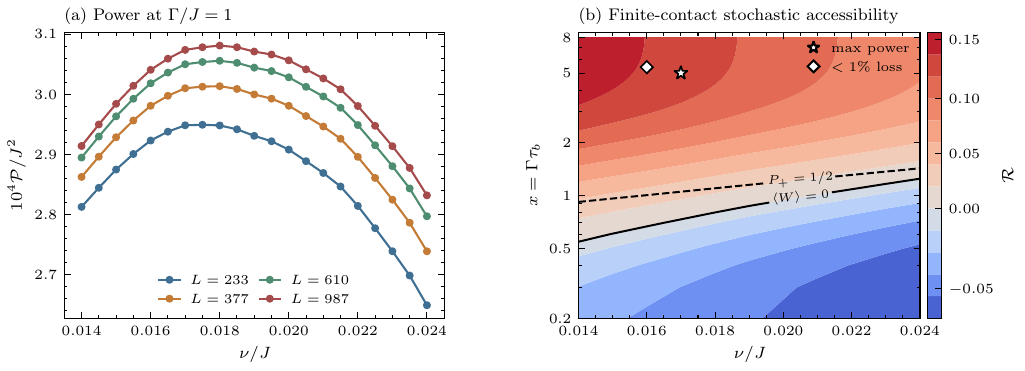}
\caption{\textbf{Finite contacts reshape mean performance and one-cycle stochastic accessibility on the crossing path.} (a) Stationary mean power versus driving rate for symmetric contacts with $\Gamma_h=\Gamma_c=J$ and $x_h=x_c=5$ at $L=233,377,610,987$. Markers are direct primary calculations and connecting lines are visual guides. (b) Signed one-cycle mean-to-width ratio $\mathcal R=\langle W\rangle/\sigma_W$ over the $(\nu/J,x)$ control plane at $L=233$, with $x=\Gamma\tau_b$. The solid contour marks $\langle W\rangle=0$ and the dashed contour $P_+=P(W>0)=1/2$. The star marks the grid maximum-power point for $\Gamma/J=1$ and the diamond a $<1\%$ power-loss compromise point.}
\label{fig:contacts}
\end{figure*}

The rapid-driving limit determines how the stroke-time cost competes with output. The smooth path ending at $r_c=0.5$ remains engine-like in the state-frozen limit, whereas the crossing path becomes non-engine-like; the latter combines a fast-drive loss of engine operation with the time cost of very slow ramps. Figure~\ref{fig:contacts}(a) shows the resulting broad maximum-power region. On the evaluated rate grid with $\Gamma_h=\Gamma_c=J$, the grid maximum lies around $\nu/J\simeq0.018$ across the four sizes (the tabulated $L=233$ point is $\nu/J=0.017$), while the peak power on the grid changes by only a few percent from $L=233$ to $987$. Faster baths shift the optimum toward larger $\nu$, but yield only modest additional power once the work-stroke duration dominates the cycle.

Mean power is only one operational criterion. Figure~\ref{fig:contacts}(b) shows how the one-cycle signed
mean-to-width ratio $\mathcal R=\langle W\rangle/\sigma_W$ varies with
driving rate and contact exposure. The mean-work and majority-positive contours remain distinct, and incomplete thermalization can reverse their ordering relative to the complete-reset case. In particular, a narrow region appears in which the stationary mean work is already positive while $P_+<1/2$: mean engine operation can be recovered before a majority of individual cycles deliver positive work.

The star and diamond in Fig.~\ref{fig:contacts}(b) show that the grid maximum-power point and the best stochastic-quality point occur at different locations. Table~\ref{tab:finite_contact_tradeoff} compares that point with a point chosen inside the $1\%$ maximum-power band on the same grid. A small reduction in power buys larger work per cycle, efficiency, and one-cycle reliability, while $P_+$ also increases. The finite-time engine has a broad operating compromise, with different optima on the evaluated grid for power, work per cycle, efficiency, and fluctuation measures, consistent with the broader power--precision tradeoffs of stochastic heat engines~\cite{Pietzonka2018,MillerTUR2021}.

\begin{table}[t]
\caption{\textbf{Representative $L=233$, $\Gamma/J=1$ crossing-path operating points.} The compromise point has the largest evaluated $\mathcal R$ within the $1\%$ maximum-power band. Here $\dot\Sigma_{\rm TPM}=\Sigma_{\rm TPM}/\tau_{\rm cyc}$.}
\label{tab:finite_contact_tradeoff}
\squeezetable
\begin{ruledtabular}
\begin{tabular}{lccc}
 & Max. power & Compromise & Weak contact \\
\hline
$\nu/J$ & 0.017 & 0.016 & 0.017 \\
$x$ & 5.0 & 5.4 & 0.8 \\
$W/J$ & 0.0286 & 0.0302 & 0.0012 \\
$\eta$ & 0.126 & 0.131 & 0.014 \\
$10^4\mathcal P/J^2$ & 2.948 & 2.921 & 0.133 \\
$\mathcal R$ & 0.130 & 0.138 & 0.006 \\
$P_+$ & 0.558 & 0.561 & 0.489 \\
$\dot\Sigma_{\rm TPM}$ & 0.0169 & 0.0159 & 0.0077 \\
\end{tabular}
\end{ruledtabular}
\end{table}

For the TPM-resolved stationary trajectory, the mean entropy
production is
\(\Sigma_{\rm TPM}=-\beta_hQ_h-\beta_cQ_c\).
It contains contributions from population relaxation during the
contacts and from the measurement-induced dephasing associated
with the TPM work protocol. Their decomposition is given in Appendix~\ref{app:tilted}; an unmeasured coherent cycle carries a different entropy accounting because the TPM dephasing step is absent.

The entropy-production rate decreases toward weak contacts, while useful output requires sufficient renewal. Table~\ref{tab:finite_contact_tradeoff} shows that reducing $\dot\Sigma_{\rm TPM}=\Sigma_{\rm TPM}/\tau_{\cyc}$ can coincide with a collapse of work, efficiency, power, and reliability. Irreversibility forms one coordinate of the operational trade-off alongside output and fluctuation measures. Finite contacts also retain information from one cycle to the next, introducing a temporal correlation scale beyond stationary one-cycle fluctuations.

\section{Repeated cycles and long-time stochastic thermodynamics}
\label{sec:longtime}

Finite contacts correlate successive cycles, so statistics accumulated over many repetitions contain information beyond a single stationary-cycle distribution~\cite{Watanabe2017,Garcez2026}. The population left at the end of one cycle becomes the input to the next, so successive work outcomes are no longer independent. The relevant question is whether the one-cycle reliability remains representative of repeated operation or whether intercycle memory changes the accumulated variance appreciably.

Full counting statistics captures these intercycle correlations within the same stationary cycle map~\cite{Esposito2009,Watanabe2017,MohantaAgarwalla2023}. For a work counting field \(\chi\), let \(\mathbb T_\chi\) denote the one-cycle population propagator with each trajectory weighted by its work increment. The scaled cumulant-generating function (SCGF) is
\begin{equation}
\psi(\chi)=\lim_{N\to\infty}\frac{1}{N}
\ln\left\langle e^{\chi X_N}\right\rangle
=\ln\rho_{\max}(\mathbb T_\chi),
\label{eq:scgf}
\end{equation}
where \(X_N\) is the accumulated work after \(N\) cycles. Derivatives of \(\psi\) at \(\chi=0\) give the long-time cumulants per cycle. In particular, for the stationary process the asymptotic work variance can be written as
\begin{equation}
\sigma_{\infty}^{2}
=
\sigma_{1}^{2}
+
2\sum_{\ell\geq1}
\operatorname{Cov}(W_q,W_{q+\ell}),
\label{eq:memoryvariance}
\end{equation}
so the difference between \(\sigma_{\infty}^{2}\) and the one-cycle variance \(\sigma_1^2\) directly measures the accumulated effect of intercycle correlations. The quantity $\mathcal R_\infty=\langle W\rangle/\sigma_\infty$ is an asymptotic reliability prefactor: for accumulated work $X_N$, $\langle X_N\rangle/\sqrt{\operatorname{Var}(X_N)}\sim\sqrt N\,\mathcal R_\infty$, rather than tending to $\mathcal R_\infty$ itself.

For the \(L=233\) crossing protocol, Table~\ref{tab:longtime_memory} shows that this correction depends strongly on the operating regime. At the refined maximum-power point, the long-time variance per cycle differs from the one-cycle value by less than \(0.3\%\). The corresponding reliability prefactor changes only from \(\mathcal R_1\simeq0.1300\) to \(\mathcal R_\infty\simeq0.1302\). Thus, at the maximum-power strong-contact point, single-cycle stationary statistics reproduce the long-time variance and reliability prefactor at the sub-percent level.

Close to the finite-contact threshold, intercycle memory becomes quantitatively important. The variance correction grows to about \(-7.8\%\), while the reliability prefactor changes from approximately \(3.70\times10^{-4}\) to \(3.86\times10^{-4}\). The negative correction means that the net sum of intercycle work covariances is negative. In this sense the repeated-cycle memory acts as an anticorrelation that suppresses the accumulated variance. This anticorrelation slightly improves the long-time reliability prefactor relative to the one-cycle estimate, while the near-threshold mean output and both reliability measures remain very small.

\begin{table}[t]
\caption{\textbf{One-cycle and long-time work reliability.} Comparison of the maximum-power point \((\nu/J,x)=(0.017,5.00)\) with a near-threshold point \((0.030,1.83)\). Here \(\mathcal R_1=\langle W\rangle/\sigma_1\), \(\mathcal R_\infty=\langle W\rangle/\sigma_\infty\), and the final row gives the relative intercycle correction to the work variance per cycle.}
\label{tab:longtime_memory}
\squeezetable
\begin{ruledtabular}
\begin{tabular}{lcc}
 & Max. power & Near threshold \\
\hline
\(\nu/J\) & 0.017 & 0.030 \\
\(x\) & 5.00 & 1.83 \\
\(\mathcal R_1\) & 0.1300 & \(3.70\times10^{-4}\) \\
\(\mathcal R_\infty\) & 0.1302 & \(3.86\times10^{-4}\) \\
\(\Delta\sigma^2/\sigma_1^2\) & \(-0.28\%\) & \(-7.78\%\) \\
\end{tabular}
\end{ruledtabular}
\end{table}

The same tilted-cycle construction can count hot and cold heat together with work. For the TPM-resolved population process defined by Eq.~\eqref{eq:contact}, the Gibbs-renewal contacts obey local detailed balance and the retraced work strokes satisfy $P_R=P_F^{\mathsf T}$; over the evaluated counting-field range, the entropy generating function displays the Gallavotti--Cohen symmetry and the joint SCGF satisfies the cycle first-law identity to numerical precision. The coherence-retaining contact channel is treated separately and is outside this monitored population construction. In the useful strong-contact regime, the one-cycle variance already approximates the long-time variance closely; near the contact threshold, the long-time statistics resolve the remaining cycle-to-cycle memory.

\section{Discussion and conclusion}

The finite-rate Otto cycle resolves a hierarchy of dynamical information
that is hidden in the usual slow-driving and state-frozen limits.
Along the quasiperiodic crossing path, thermodynamic recovery does not
coincide with microscopic energy-ordered following. Positive mean work
returns while the transition matrix remains strongly redistributed,
because population transferred away from the energy-ordered channels is
progressively redirected toward energetically less costly processes.
Across the recovery region, the total amount of population redistribution
changes only weakly while its energetic penalty decreases substantially.
The relevant distinction is therefore not simply how much population is
transferred, but how much energetic weight those transitions carry.

The spectral analysis identifies where this redistribution is generated.
Thermally occupied regions with large spectral susceptibility carry a
disproportionate share of the excess-energy growth, while the
complete-cycle deficit remains distributed over a broad set of spectral
regions. The narrowest isolated avoided crossing accounts for only a
small fraction of the deficit in the representative recovery regime.
The equal-duration schedule control provides an independent dynamical
test: reallocating time toward high-susceptibility regions improves the
complete-cycle work, whereas reallocating the same time toward
low-susceptibility regions worsens it, even though the transition map
remains far from energy-ordered following. Spectral accessibility is
therefore determined jointly by dynamical coupling, thermal occupation,
and energetic weight rather than by a single spectral scale.

The work distribution resolves a further separation of dynamical scales.
The majority-positive and positive-mean boundaries remain distinct over
a finite rate interval for all four Fibonacci sizes studied. Within the
negative-work sector, the conditional energetic loss increases strongly
with branch displacement and is concentrated in the thermally occupied
low-energy manifold. Mean operation, majority-cycle success,
fluctuation-based reliability, and microscopic energy-ordered following
consequently probe different coarse grainings of the same finite-rate
unitary transformation. Table~\ref{tab:physics_summary} summarizes this
hierarchy together with the additional information introduced by finite
thermal contacts and repeated operation.

\begin{table}[t]
\caption{\textbf{Hierarchy of finite-time thermodynamic information.}
Each diagnostic resolves a different layer of the same driven
quasiperiodic cycle.}
\label{tab:physics_summary}
\begin{ruledtabular}
\begin{tabular}{@{}ll@{}}
\parbox[t]{0.30\columnwidth}{\raggedright Diagnostic layer} &
\parbox[t]{0.62\columnwidth}{\raggedright Physical content} \\
\hline
\parbox[t]{0.30\columnwidth}{\raggedright Finite-rate accessibility} &
\parbox[t]{0.62\columnwidth}{\raggedright
The response is distributed across thermally occupied, dynamically
coupled spectral regions. Equal-duration time reallocation guided by the
spectral susceptibility changes the work systematically, while the
narrowest isolated crossing contributes only a small fraction of the
complete-cycle deficit.} \\[2pt]
\parbox[t]{0.30\columnwidth}{\raggedright Mean work and branch following} &
\parbox[t]{0.62\columnwidth}{\raggedright
Positive mean work can recover while the transition map remains strongly
redistributed because the surviving off-diagonal transfer becomes
energetically less costly.} \\[2pt]
\parbox[t]{0.30\columnwidth}{\raggedright Fluctuation resolution} &
\parbox[t]{0.62\columnwidth}{\raggedright
The majority-positive and positive-mean boundaries remain separated
across the four sizes, while large-displacement negative-work
trajectories carry systematically larger losses.} \\[2pt]
\parbox[t]{0.30\columnwidth}{\raggedright Finite thermal contacts} &
\parbox[t]{0.62\columnwidth}{\raggedright
Within the specified Gibbs-renewal model, incomplete population renewal
creates a second accessibility constraint and separates the optima found
on the evaluated grid for work, power, efficiency, reliability, and
irreversibility.} \\[2pt]
\parbox[t]{0.30\columnwidth}{\raggedright Repeated operation} &
\parbox[t]{0.62\columnwidth}{\raggedright
Strong-contact operation is well represented by one-cycle variance,
while contact-threshold operation develops appreciable intercycle
memory.} \\
\end{tabular}
\end{ruledtabular}
\end{table}

The quasiperiodic structure determines the rate and path dependence of
this hierarchy. Matched-temperature, matched-duration, and
equal-excursion controls retain the enhanced finite-rate sensitivity of
the crossing route, while changes in modulation, phase, and
diagonal--off-diagonal balance shift the accessibility boundary by
reshaping the spectral path and endpoint thermodynamics. Cooling
restricts the active spectrum toward the low-energy sector, where the
finite-rate response develops reproducible Kibble--Zurek-like scaling
over finite windows. At the opposite end of the rate axis, the
state-frozen reference develops pronounced Loschmidt-type structure,
whereas the continuous ramps interpolate smoothly toward that limit.
The fluctuation ordering is likewise protocol dependent: the localised
control is broadest in the strict slow-driving reference, while the
critical control dominates in the state-frozen limit and the finite-rate
maximum shifts toward the critical--localised crossover.

Finite thermal contacts add an independent dynamical timescale.
Within the Gibbs-renewal model, incomplete population renewal can
suppress work extraction even when the isolated strokes already lie in
the positive-work regime. The coherence-retaining control further shows
that weak-contact operation can depend on the dynamical channel.
Because work-stroke and contact durations both enter the cycle time,
finite-contact operation produces distinct compromises among work per
cycle, power, efficiency, reliability, and irreversibility. The same
incomplete renewal generates intercycle correlations. These correlations
remain weak near the useful maximum-power region considered here but
become appreciable near the contact threshold, where long-time counting
statistics are needed to characterize the fluctuations consistently.

Taken together, these results establish a finite-rate thermodynamic
picture of quasiperiodic spectral accessibility. The cycle responds to
a thermally weighted network of dynamical pathways whose energetic
relevance depends on the driving protocol, rather than to a single
avoided crossing or static localisation label. Mean work, trajectory
statistics, microscopic transition maps, and long-time fluctuations then
resolve successive levels of the same structured dynamics. Thermodynamic
recovery, stochastic success, and microscopic following are consequently
distinct physical notions, and their separation provides an operational
route for resolving finite-rate spectral structure through quantum
thermodynamic observables.

\noindent\textit{Outlook.}
The present work establishes a controlled single-particle setting in
which finite-rate spectral accessibility can be followed from
microscopic population transfer to mean thermodynamics and full work
statistics. A first research program is to determine which aspects of
this hierarchy are specifically enhanced by quasiperiodic order and which
belong more generally to structured multilevel dynamics. This requires
matched periodic, disordered, and engineered controls in which the
thermally active bandwidth, gap distribution, path length, and driving
duration are placed on comparable footing. Such a comparison would turn
the present AAH result into a broader test of the universality of
spectral-accessibility mechanisms. Introducing finite filling and
interactions provides the next stage of the same program. Occupation
constraints, level repulsion, scattering, and many-body level structure
would then allow the separation between microscopic following, positive
mean operation, and stochastic reliability to be tested when the
accessible spectrum is generated collectively rather than by a
single-particle band problem. These questions require new Hamiltonian
families and many-body dynamical calculations beyond the controlled
single-particle setting considered here.

A second research program concerns the reservoir and coherence sectors
of the finite-rate mechanism. The Gibbs-renewal contacts used here
isolate the work-stroke dynamics and provide a transparent two-clock
description. Microscopic reservoirs would instead introduce
energy-dependent transition rates, eigenstate-selective couplings, and
bath-induced coherence or dephasing, allowing the spectral structure of
the working medium and that of the environment to be treated on the same
footing. The coherence-retaining control already shows that weak-contact
operation can depend on the dynamical channel, motivating a direct
comparison of monitored and coherence-preserving fluctuation
statistics~\cite{Son2021Monitoring}. Within the same program, the
gauge-invariant formulation of finite-time Otto thermodynamics developed
in Ref.~\cite{Sambu2026Gauge} is directly relevant to the central
separation found here: positive mean work recovers while energy-ordered
microscopic following remains strongly broken. Applying this
decomposition to the driven AAH strokes could determine whether that
recovery is carried predominantly by the sector associated with the
instantaneous spectrum and populations, by coherence generated during
finite-rate driving, or by a cancellation between the two contributions.
This would distinguish thermodynamic recovery arising from the energetic
reweighting of strongly redistributed populations from recovery that
depends essentially on quantum coherence. Addressing this question
requires access to the full density matrix throughout the driven strokes,
rather than only the population transition maps sufficient for the
present analysis, and therefore defines a dedicated coherence-resolved
study.

A third research program is to use the finite-rate cycle as an inference
tool for spectral structure. The present results show that temperature,
driving rate, and path select different dynamically active regions of
the spectrum and leave distinct signatures in work and heat statistics.
Scanning these controls systematically could provide an overconstrained
thermodynamic data set from which properties of the underlying spectral
dynamics may be inferred without reconstructing the complete microscopic
state. Establishing such an inverse protocol requires combining many
operating points and determining which spectral information is uniquely
encoded in the thermodynamic response. The low-temperature regime offers
a complementary route toward asymptotic critical dynamics. Larger
systems, slower ramps, and scaling windows stable against the
quasiperiodic gap hierarchy would allow the finite-window
Kibble--Zurek-like behavior found here to be connected to a controlled
asymptotic scaling theory. These directions extend the same central
question---how structured quantum dynamics becomes visible through
thermodynamic observables---toward universality, many-body physics,
microscopic reservoirs, coherence-resolved energetics, and spectral
inference.

\section*{Author contributions}
S.C. contributed to the methodology, numerical investigation, and
software implementation. L.C.C., N.G.A., and G.X. contributed to the
conceptualization and validation of the work. G.D.M.N. contributed to
the conceptualization, methodology, theoretical analysis, numerical
investigation, visualization, and preparation of the original draft.
All authors discussed the results, reviewed the manuscript, and
approved the submitted version. OpenAI ChatGPT (GPT-5.6 Sol) was used
for language editing, bibliographic organization, and code-assisted
analytical and numerical cross-checks. The authors specified the physical
model and validation criteria, checked the resulting calculations against
analytic identities and numerical convergence tests, and assume full
responsibility for the manuscript.

\begin{acknowledgments}
	This research is supported by Zhejiang Provincial Natural Science Foundation of China under Grant No. LQN25A040012, the start-up fund from Xingzhi College, Zhejiang Normal University, and the National Natural Science Foundation of China under Grant No. 12604416. LCC acknowledges support from the Brazilian National Council for Scientific and Technological Development (CNPq) through grant 305636/2026-0, the National Institute of Science and Technology for Applied Quantum Computing through CNPq grant 408884/2024-0, Goiás Research Foundation (FAPEG) through grant 202510267001843, and São Paulo Research Foundation (FAPESP) through grant 2025/23726-4.
\end{acknowledgments}

\section*{Data and code availability}
Data and numerical code supporting the findings of this study are available from the corresponding author upon reasonable request.

\bibliography{references}

\appendix
\section{Finite-rate propagation and transition diagnostics}
\label{app:propagation}

For a linear ramp of duration \(\tau=|r_c-r_h|/\nu\), the interval is divided into \(N=\lceil\tau/\Delta\tau_{\max}\rceil\) equal steps of size \(h=\tau/N\). The Hamiltonian is evaluated at each step's midpoint. For \(\hbar=1\), the midpoint Cayley update is
\begin{equation}
	\left(I+\frac{i h H_{j+1/2}}{2}\right)\ket{\psi_{j+1}}
	=\left(I-\frac{i h H_{j+1/2}}{2}\right)\ket{\psi_j}.
	\label{eq:cayley}
\end{equation}
This second-order update is unitary up to the linear-solver tolerance. Choosing \(N\) from the ceiling function keeps the final control parameter exactly at the requested endpoint even when \(\tau/\Delta\tau_{\max}\) is not an integer.

The crossing-path transition and work anchors use the dimensionless timestep $\Delta\tau_{\max}=0.0025$ and agree with a full-basis adaptive DOP853 reference with relative and absolute tolerances $2\times10^{-11}$ and $2\times10^{-13}$, respectively. The stochastic maps in Fig.~\ref{fig:work} use $\Delta\tau_{\max}=0.005$. Relative to the finer $0.0025$ grid, this changes the mean work by at most $6.6\times10^{-6}J$ over the common evaluated plane and $P(W>0)$ by at most $1.3\times10^{-6}$ at the three representative rates. The finite-contact power scans also use $\Delta\tau_{\max}=0.005$, which sets their quoted numerical resolution.

For real \(H(r)\) and a reversed protocol, the forward and reverse transition matrices obey \(P_R=P_F^{\mathsf T}\). The propagated matrices satisfy row and column normalization and the transpose relation within numerical precision. The normalized distances from the energy-ordered and state-frozen reference maps are
\begin{align}
	D_{\ad}&=\frac{\|P_F-I\|_F}{\sqrt L},\nonumber\\
	D_{\sud}&=\frac{\|P_F-P_{\sud}\|_F}{\sqrt L},\nonumber\\
	P_{\sud}(m|n)&=|\langle m_c|n_h\rangle|^2 .
	\label{eq:transitiondistances}
\end{align}
These diagnostics compare population transition matrices; phase-sensitive information resides in the full propagator. For orientation, a uniformly mixing matrix gives $D_{\ad}=\sqrt{1-1/L}$, so values near unity diagnose diffuse population transfer. The associated energy scale is supplied separately by the Gibbs-weighted leakage and the energy-weighted work deficit. For an initial Gibbs distribution \(p_n\), the probability weight transferred away from its energy-ordered channel is \(\ell=\sum_n p_n[1-P_F(n|n)]\); the reverse leakage is defined with \(P_R\) and the cold Gibbs distribution.

The cycle-level work deficit \(W_{\fric}=W_{\ad}-W_\nu\) compares the work output of the strict energy-ordered reference with that of a finite-rate cycle and serves as the cycle-level measure of the finite-rate work penalty.

\subsection{Transition-resolved work deficit}
\label{app:transition_work}

The work deficit can be resolved directly into the transitions generated during the two isolated strokes. For the forward stroke,
\begin{equation}
\Delta W_F
=
\sum_{n,m}
p_{hn}P_F(m|n)
\left(E_m^c-E_n^c\right),
\label{eq:transition_deficit_forward}
\end{equation}
while the reverse stroke contributes
\begin{equation}
\Delta W_R
=
\sum_{k,\ell}
p_{ck}P_R(\ell|k)
\left(E_\ell^h-E_k^h\right).
\label{eq:transition_deficit_reverse}
\end{equation}
Their sum gives
\begin{equation}
W_{\ad}-W_\nu=\Delta W_F+\Delta W_R .
\label{eq:transition_deficit_total}
\end{equation}
The deficit is nonnegative under the assumptions of the complete-reset protocol. For either stroke, order the positive-temperature Gibbs populations as $p_1\geq p_2\geq\cdots\geq p_L$ and the endpoint energies as $E_1\leq E_2\leq\cdots\leq E_L$. A unitary transition matrix is unistochastic and hence doubly stochastic, so by the Birkhoff--von Neumann theorem it is a convex combination of permutation matrices. The rearrangement inequality then gives
\begin{equation}
\bm E^{\mathsf T}P\bm p\geq\bm E^{\mathsf T}\bm p,
\label{eq:passivity_rearrangement}
\end{equation}
because the energy-ordered assignment pairs the largest populations with the lowest endpoint energies. Applying this result to the two strokes separately yields
\begin{equation}
\Delta W_F\geq0,\qquad \Delta W_R\geq0,\qquad
W_{\ad}-W_\nu\geq0.
\label{eq:work_deficit_nonnegative}
\end{equation}
The proof assumes positive-temperature Gibbs populations; degeneracies make the inequalities non-strict. The diagonal terms in Eqs.~\eqref{eq:transition_deficit_forward} and~\eqref{eq:transition_deficit_reverse} vanish identically. At fixed endpoints and bath temperatures, the Gibbs weights and endpoint energy differences are independent of the driving rate, so the rate dependence of the work deficit comes entirely from the redistribution of transition probabilities.

Table~\ref{tab:transition_work_decomposition} shows the decomposition for the crossing protocol at $L=233$, $T_h/J=0.65$, and $T_c/J=0.1$. Across the mean-work recovery boundary, from $\nu/J=0.04$ to $0.03$, the total deficit decreases by $23.69\%$, whereas the mean thermally weighted leakage decreases by only $0.185\%$. Defining
\begin{equation}
\epsilon_{\rm leak}
=
\frac{W_{\ad}-W_\nu}{\ell_F+\ell_R},
\label{eq:energetic_penalty_leak}
\end{equation}
this net work-deficit-to-leakage ratio falls by $23.55\%$. The probability of leaving the energy-ordered channels remains nearly unchanged while the net work deficit associated with that redistributed population decreases substantially.

The reverse stroke accounts for $81.45\%$ of the reduction in the total deficit between the two rates. Grouping transition changes by branch displacement, with $d=|m-n|$ on the forward stroke and $d=|\ell-k|$ on the reverse stroke, the sector $d\geq20$ accounts for $98.16\%$ of the net recovery; smaller-displacement changes partially oppose it. The recovery is distributed over many transitions: the twenty largest positive transition-level changes contain only $4.30\%$ of the total positive recovery weight. The mean work recovers through a collective shift of the off-diagonal transition map toward channels carrying smaller net work penalties.

\setcounter{table}{0}
\renewcommand{\thetable}{A\arabic{table}}
\renewcommand{\theHtable}{A\arabic{table}}

\begin{table*}[t]
\caption{\textbf{Transition-resolved origin of mean-work recovery on the crossing path.}
Rate-resolved work deficit and microscopic population redistribution for $L=233$, $T_h/J=0.65$, and $T_c/J=0.1$. The lower block resolves the change from $\nu/J=0.04$ to $0.03$. Here $\epsilon_{\rm leak}=(W_{\ad}-W_\nu)/(\ell_F+\ell_R)$ is the signed net work-deficit-to-leakage ratio and includes cancellations among transition contributions.}
\label{tab:transition_work_decomposition}
\squeezetable
\begin{ruledtabular}
\begin{tabular}{lcccccc}
$\nu/J$
& $\langle W\rangle/J$
& $(W_{\ad}-W_\nu)/J$
& $\Delta W_F/J$
& $\Delta W_R/J$
& $\bar\ell$
& $D_{\ad}$ \\
\hline
0.040 & $-6.98\times10^{-4}$ & 0.05289 & 0.00983 & 0.04307 & 0.96765 & 0.98285 \\
0.030 & 0.01183 & 0.04036 & 0.00750 & 0.03286 & 0.96585 & 0.98249 \\
0.017 & 0.02928 & 0.02292 & 0.00432 & 0.01860 & 0.96171 & 0.98170 \\
\hline
\multicolumn{4}{l}{Reduction of total work deficit} &
\multicolumn{3}{c}{23.69\%} \\
\multicolumn{4}{l}{Reduction of mean leakage} &
\multicolumn{3}{c}{0.185\%} \\
\multicolumn{4}{l}{Reduction of $\epsilon_{\rm leak}$} &
\multicolumn{3}{c}{23.55\%} \\
\multicolumn{4}{l}{Reverse-stroke share of deficit reduction} &
\multicolumn{3}{c}{81.45\%} \\
\multicolumn{4}{l}{Net recovery from $d\geq20$} &
\multicolumn{3}{c}{98.16\%} \\
\multicolumn{4}{l}{Largest 20 transitions / positive recovery weight} &
\multicolumn{3}{c}{4.30\%} \\
\end{tabular}
\end{ruledtabular}
\end{table*}

\section{Finite-rate and finite-size robustness}
\label{app:accessibility_support}

\setcounter{figure}{0}
\renewcommand{\thefigure}{B\arabic{figure}}
\renewcommand{\theHfigure}{B\arabic{figure}}
\setcounter{table}{0}
\renewcommand{\thetable}{B\arabic{table}}
\renewcommand{\theHtable}{B\arabic{table}}

This appendix collects the rate- and size-resolved diagnostics supporting the accessibility analysis. It locates the mean-work and engine-window boundaries, follows the thermally relevant low-spacing hierarchy and transition dynamics through \(L=987\), and records the numerical convergence scale. The propagation errors remain well below the finite-rate variations reported here.

\begin{table*}[t]
	\caption{\textbf{Numerical landmarks for the crossing protocol.} Representative values locating the finite-rate operating boundaries, their finite-size shift, and the associated branch exchange.}
	\label{tab:app_certification}
	\begin{ruledtabular}
		\begin{tabular}{p{0.26\textwidth} p{0.25\textwidth} p{0.41\textwidth}}
			Numerical question & Condition & Numerical result \\
			\hline
			Strict-reference engine interval
			& crossing path, \(L=233\), \(T_c/J=0.1\)
			& \(0.212<T_h/J<1.403\) \\
			Finite-rate engine interval
			& same path, \(\nu/J=0.03\)
			& \(0.454<T_h/J<0.974\) \\
			Mean-work sign change
			& \(L=233\), \(T_h/J=0.65\)
			& \(0.0394<\nu/J<0.0397\) \\
			Disappearance of any engine window
			& \(L=233\), scan over \(T_h\)
			& \(0.0397<\nu/J<0.0398\) \\
			Finite-size shift of the work boundary
			& \(\nu/J=0.04\)
			& \(\langle W\rangle/J=-6.99\times10^{-4}\) (\(L=233\)),
			\(+1.9\times10^{-5}\) (377), \(+7.66\times10^{-4}\) (987) \\
			Positive-work probability near that boundary
			& \(\nu/J=0.04\)
			& \(P(W>0)=0.506\) (\(L=233\)) and \(0.508\) (987) \\
			Resolved branch-character exchange
			& \(L=89\), \(r_*\simeq1.044\)
			& same/exchanged overlap weights change from \(0.844/0.156\) at \(\delta r=10^{-4}\)
			to \(0.179/0.821\) at \(\delta r=5\times10^{-4}\) \\
		\end{tabular}
	\end{ruledtabular}
\end{table*}

\begin{figure}[t]
	\centering
	\includegraphics[width=\columnwidth]{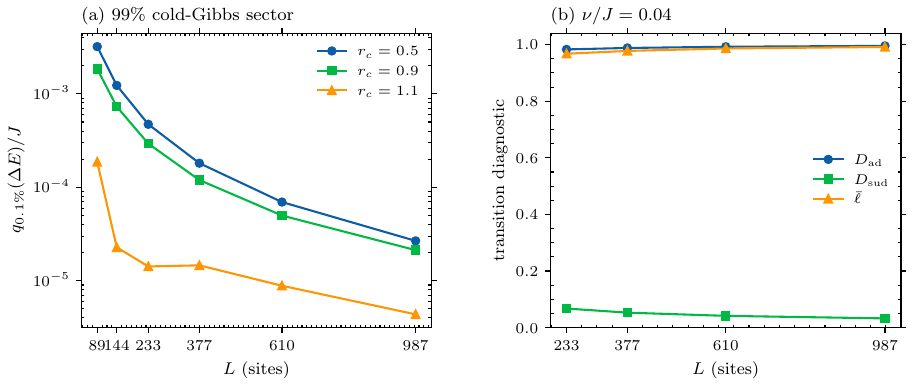}
	\caption{\textbf{Finite-size robustness of the spectral and transition structure.}
		(a) Low-spacing tail within the lowest energy-ordered sector carrying \(99\%\)
		of the cold-endpoint Gibbs probability. Adjacent spacings are collected from
		101 uniformly spaced Hamiltonians along each work-stroke path and the \(0.1\%\)
		quantile is plotted. The crossing path remains separated from both controls
		through \(L=987\).
		(b) Microscopic transition diagnostics at \(\nu/J=0.04\) for
		\(L=233,377,610,987\). The finite-rate population map remains far from
		energy-ordered following while staying close to the state-frozen population map,
		with near-unit Gibbs-weighted leakage. Together with the \(\nu/J=0.03\) values in
		Table~\ref{tab:size_persistence}, these data show persistent microscopic
		nonadiabaticity across the thermodynamic recovery region.}
	\label{fig:app_numerical_certification}
\end{figure}

Figure~\ref{fig:app_numerical_certification}(a) follows the large-\(L\) spectral hierarchy with a low-spacing quantile that remains above the numerical conditioning floor. For each size and path, adjacent instantaneous spacings are pooled from 101 uniformly spaced Hamiltonians along the work stroke. The plotted \(0.1\%\) quantile is restricted to the lowest energy-ordered levels carrying \(99\%\) of the cold-endpoint Gibbs probability at \(T_c/J=0.1\). The crossing path retains a substantially lower spacing tail than both controls through \(L=987\). At that size the quantile is smaller by factors of approximately \(6.1\) and \(4.9\) relative to the \(r_c=0.5\) and \(r_c=0.9\) controls, respectively. The same ordering is obtained from the full spectrum, confirming the low-spacing
hierarchy in both the thermally restricted and full spectral data.

The complementary \(\nu/J=0.04\) size dependence is shown in
Fig.~\ref{fig:app_numerical_certification}(b), where the precise mean-work boundary is
already finite-size dependent in Table~\ref{tab:app_certification}. Nevertheless,
\(D_{\ad}\) is close to unity, \(D_{\sud}\) is small, and the mean
Gibbs-weighted leakage \(\bar\ell\) is near unity across
\(L=233,377,610,987\). In particular,
\(D_{\ad}:0.983\to0.995\),
\(D_{\sud}:0.0679\to0.0331\), and
\(\bar\ell:0.968\to0.992\) between the smallest and largest displayed sizes.
The corresponding \(\nu/J=0.03\) four-size values are listed in main
Table~\ref{tab:size_persistence}; the Appendix uses the complementary rate to provide an independent view of the same size sequence.

The propagation and reciprocity convergence data for these calculations are collected in
Appendix~\ref{app:propagation}. At \(L=987\), reducing the dimensionless timestep \(\Delta\tau_{\max}\) from \(0.005\) to
\(0.0025\) changes the work only at the \(7\times10^{-7}\)--\(8\times10^{-7}J\)
scale at the two rates used in the size calculations, well below the finite-rate changes
reported in the main text.

\begin{table*}[t]
\caption{\textbf{Finite-size separation of the mean-work and majority-positive boundaries.}
All values use the crossing path at $T_h/J=0.65$ and $T_c/J=0.1$. The two middle columns give the numerical brackets containing the zeros of $\langle W\rangle$ and $P(W>0)-1/2$. The final column is the interval width obtained by linear interpolation within those brackets. At the common rate $\nu/J=0.041$, all four sizes lie inside the sign-separated interval.}
\label{tab:size_sign_boundaries}
\squeezetable
\begin{ruledtabular}
\begin{tabular}{lccccc}
$L$ & $\langle W\rangle/J$ at $0.041$ & $P_+$ at $0.041$ &
$\langle W\rangle=0$ bracket & $P_+=1/2$ bracket & $\Delta\nu_{\rm sep}/J$ \\
\hline
233 & $-1.962\times10^{-3}$ & 0.50427 & $0.039375$--$0.039688$ & $0.0425$--$0.0450$ & 0.00417 \\
377 & $-1.241\times10^{-3}$ & 0.50533 & $0.0400$--$0.0410$ & $0.0425$--$0.0450$ & 0.00429 \\
610 & $-7.705\times10^{-4}$ & 0.50602 & $0.0400$--$0.0410$ & $0.0425$--$0.0450$ & 0.00437 \\
987 & $-4.908\times10^{-4}$ & 0.50640 & $0.0400$--$0.0410$ & $0.0450$--$0.0475$ & 0.00440 \\
\end{tabular}
\end{ruledtabular}
\end{table*}

At $L=987$ and $\nu/J=0.041$, reducing the dimensionless timestep from $0.005$ to $0.0025$ changes $\langle W\rangle/J$ by $8.3\times10^{-7}$ and $P_+$ by $1.2\times10^{-6}$, well below the distances from the two sign boundaries.

\subsection{Matched controls and thermally weighted spectral response}
\label{app:matched_spectral}

To separate spectral-path sensitivity from the different representative temperatures and stroke durations, we compare the three original paths at common $T_h/J=0.65$, $T_c/J=0.1$, and common work-stroke duration $\tau=18.5167$. The results are summarised in Table~\ref{tab:matched_controls_spectral}. The crossing path remains the only work-consuming protocol and retains the largest work width, while its transition map and thermally occupied sector are much farther from energy-ordered following than the two controls.

A direct spectral-response diagnostic is obtained from
\begin{align}
g_{rr}^{(n)}(r)
&=\sum_{m\neq n}
\frac{|\langle m(r)|\partial_r H|n(r)\rangle|^2}
{[E_m(r)-E_n(r)]^2},\nonumber\\
\chi_b(r)&=J^2\sum_n p_{bn}g_{rr}^{(n)}(r),
\label{eq:spectral_susceptibility}
\end{align}
where $p_{bn}$ are the Gibbs weights at the beginning of the corresponding stroke, carried along the energy-ordered labels. Away from exact degeneracies, $g_{rr}^{(n)}$ is the band quantum metric along the driven coordinate $r$~\cite{Kolodrubetz2017}; hence $\chi_b/J^2$ is a population-weighted sum of band quantum metrics. We refer to $\chi_b$ as the \emph{thermally weighted spectral susceptibility}. A full mixed-state thermal fidelity metric also contains population-dependent terms. Equation~\eqref{eq:spectral_susceptibility} weights instantaneous interlevel couplings by Gibbs occupation and inverse squared spacing; finite-rate transition probabilities additionally depend on dynamical phases and interference.

For the crossing path, $\log\chi_b$ is strongly rank-correlated with the local excess-energy increments. At $\nu/J=0.04$, the Spearman rank coefficients between $\log\chi_b$ and the absolute excess-energy increments are $0.842$ on the forward stroke and $0.650$ on the reverse stroke. The $20\%$ of forward-stroke intervals with the largest $\chi_b$ carry $62.8\%$ of the positive excess-energy growth, and the largest $30\%$ carry $76.6\%$. At the same stroke duration, the maximum forward susceptibility increases from approximately $1.0$ on the extended path and $11.6$ on the near-critical path to $6.4\times10^3$ on the crossing path; the reverse-stroke maximum reaches $8.2\times10^3$.

To resolve the complete-cycle energetic contribution spatially, define the instantaneous forward excess energy
\begin{equation}
\delta E_F(r)=\Tr[\rho_F(r)H(r)]-\sum_n p_{hn}E_n(r),
\label{eq:regional_excess_forward}
\end{equation}
and analogously $\delta E_R(r)$ with the cold Gibbs weights on the reverse stroke. For spatial bins $[r_j,r_{j+1}]$, the signed increments $C_j^F$ and $C_j^R$ are the corresponding changes in $\delta E_F$ and $\delta E_R$, with the reverse bins mapped onto the same ascending-$r$ grid. They satisfy exactly
\begin{equation}
\sum_j(C_j^F+C_j^R)=\Delta W_F+\Delta W_R.
\label{eq:regional_deficit_sum}
\end{equation}
At $\nu/J=0.04$, ranking bins by $\chi_h+\chi_c$ shows that the top $10\%$, $20\%$, and $30\%$ of the path carry $15.88\%$, $31.64\%$, and $49.96\%$ of the signed cycle deficit. Reaching $50\%$, $80\%$, and $90\%$ of the positive cycle deficit requires 29, 59, and 78 of 100 spatial bins, respectively. The reverse-stroke contribution is broader than the forward one and dominates the total deficit.

A refined two-level fit near $r/J\simeq1.041003$ gives a minimum gap $\Delta/J\simeq4.81\times10^{-8}$ and Landau--Zener exponents of order $10^{-11}$ at both $\nu/J=0.04$ and $0.03$, so this pair is effectively diabatic at both rates. Under the additive regional definition above, the window $|r-r_*|/J\leq0.005$ contributes $4.83\%$ and $5.19\%$ of the forward deficits at the two rates, but only $2.01\%$ and $2.27\%$ of the complete-cycle deficits because the reverse contribution is both larger and more broadly distributed. The cycle-level work penalty is therefore distributed well beyond this minimum-gap pair.

The geometric diagnostic can also be tested operationally without changing the total stroke time. Divide the path into 100 equal spatial intervals and rank them by $\chi_\Sigma(r)=\chi_h(r)+\chi_c(r)$. In the high-susceptibility schedule, the top $20\%$ receive $1.25$ times the uniform interval duration and the bottom $20\%$ receive $0.75$ times that duration; the middle $60\%$ are unchanged. The low-susceptibility control reverses these allocations, and the reverse stroke retraces the same spatial schedule. Because the top and bottom sets have equal width, all three schedules have exactly the same total duration.

At $\nu/J=0.04$ ($\tau=18.5167$), the high-susceptibility schedule changes $\langle W\rangle/J$ from $-6.98\times10^{-4}$ to $+3.12\times10^{-3}$ and lowers the work deficit by $7.22\%$, whereas the opposite allocation gives $-4.79\times10^{-3}$ and raises the deficit by $7.74\%$. The ordering occurs on both work strokes. At $\nu/J=0.03$ the same control gives $0.01634$, $0.01184$, and $0.00670$ for the high-susceptibility, uniform, and low-susceptibility schedules, respectively. Throughout, $D_{\ad}\simeq0.982$--$0.983$ and the mean Gibbs-weighted leakage changes by less than $5\times10^{-4}$; the energetic improvement occurs while energy-ordered following remains essentially unchanged.

\begin{table*}[t]
\caption{\textbf{Matched-duration path control and spectral susceptibility.}
All three paths use $L=233$, $T_h/J=0.65$, $T_c/J=0.1$, and $\tau=18.5167$. The rate is adjusted for each excursion to keep the stroke duration fixed. $\chi_F^{\max}$ and $\chi_R^{\max}$ are the maxima of Eq.~\eqref{eq:spectral_susceptibility} on the forward and reverse strokes.}
\label{tab:matched_controls_spectral}
\squeezetable
\begin{ruledtabular}
\begin{tabular}{lccccccc}
Path & $\nu/J$ & $\langle W\rangle/J$ & $\sigma_W/J$ & $D_{\ad}$ & $\bar\ell$ & $\chi_F^{\max}$ & $\chi_R^{\max}$ \\
\hline
Extended & 0.00540 & 0.01303 & 0.02756 & 0.0173 & 0.00453 & 1.03 & 0.43 \\
Near-critical & 0.03273 & 0.04649 & 0.19013 & 0.3589 & 0.24050 & 11.64 & 11.51 \\
Crossing & 0.04000 & $-6.99\times10^{-4}$ & 0.25814 & 0.9828 & 0.96765 & $6.42\times10^3$ & $8.16\times10^3$ \\
\end{tabular}
\end{ruledtabular}
\end{table*}

\begin{table*}[t]
\caption{\textbf{Protocol dependence of work fluctuations and operating modes.}
All fluctuation controls use $L=233$, $T_h/J=0.65$, $T_c/J=0.1$,
and equal excursions $\Delta r/J=0.1$ unless stated otherwise.
The regional scan moves an equal-width interval across the $r$ axis.
The operating-mode classification for the crossing path uses the
temperature grid $0.15\leq T_h/J\leq1.60$.}
\label{tab:equal_excursion_fluctuations}
\begin{ruledtabular}
\begin{tabular}{llccc}
\multicolumn{5}{c}{\textbf{(a) Equal-excursion work widths}}\\
Control & Path & Slow reference & $\nu/J=0.04$ & State-frozen \\
\hline
& Extended & 0.0270 & 0.0433 & 0.1131 \\
& Critical & 0.0228 & 0.0598 & 0.1243 \\
& localised & 0.0366 & 0.0464 & 0.1056 \\
\hline
\multicolumn{5}{c}{\textbf{(b) Largest width in the sliding equal-excursion scan}}\\
Protocol & $r_h/J\to r_c/J$ & midpoint $r_m/J$ &
$\sigma_W/J$ & \\
\hline
Slow reference & $1.65\to1.75$ & 1.70 & 0.0373 & \\
$\nu/J=0.04$ & $1.00\to1.10$ & 1.05 & 0.0608 & \\
$\nu/J=0.25$ & $1.00\to1.10$ & 1.05 & 0.1179 & \\
State-frozen & $0.95\to1.05$ & 1.00 & 0.1248 & \\
\hline
\multicolumn{5}{c}{\textbf{(c) Crossing-path operating modes}}\\
Protocol & Refrigerator sector & Engine sector &
Other modes on the grid & \\
\hline
Slow reference &
$0.15\lesssim T_h/J\lesssim0.185$ &
present & accelerator, heater & \\
$\nu/J=0.03$ &
absent & present & accelerator, heater & \\
$\nu/J=0.04,0.05,0.10,0.25,0.50,1,10$ &
absent & absent & accelerator, heater & \\
State-frozen &
absent & absent & accelerator, heater & \\
\end{tabular}
\end{ruledtabular}
\end{table*}

\section{Two-projective-measurement work statistics}
\label{app:tpm}

\setcounter{table}{0}
\renewcommand{\thetable}{C\arabic{table}}
\renewcommand{\theHtable}{C\arabic{table}}

For the forward stroke, the output work \(w_F=E_n^h-E_m^c\) occurs with probability \(p_{hn}P_F(m|n)\). For the reverse stroke, \(w_R=E_k^c-E_\ell^h\) occurs with probability \(p_{ck}P_R(\ell|k)\). Complete Gibbs resets make these two stroke outcomes independent, so \(P_{\cyc}(W)\) is their convolution. The first two moments and sign-conditioned quantities can be written directly from the discrete outcomes:
\begin{align}
	\langle W\rangle&=\sum_a\pi_a W_a,&
	\sigma_W^2&=\sum_a\pi_a(W_a-\langle W\rangle)^2,\nonumber\\
	P_+&=\sum_{a:W_a>0}\pi_a,&
	\mathcal R&=\frac{\langle W\rangle}{\sigma_W},
	\label{eq:tpmstats}
\end{align}
where \(a\) labels an outcome pair with probability \(\pi_a\) and work \(W_a\). The signed ratio is retained when \(\langle W\rangle<0\); it is called engine reliability only in the positive-mean regime.

Exact sign probabilities and conditional means are evaluated from the unbinned discrete work distribution. Each stroke has \(L^2\) level-pair outcomes. We sort the reverse-stroke work values and form cumulative probability and probability-weighted-work sums. For each forward outcome, a binary search identifies the reverse outcomes for which \(w_R>-w_F\). Prefix sums then give \(P(W>0)\), the positive-work sum, and the complementary nonpositive-work sum with \(O(L^2\log L)\) sorting and search work after the transition matrices have been computed. The conditional means are the exact sign-restricted work sums divided by their probabilities. Any binned density is only a visualization of the discrete convolution; all quoted moments and sign probabilities use the unbinned outcomes.

The sign-separated regime remains visible after excluding near-zero outcomes. At $L=233$, $T_h/J=0.65$, and $\nu/J=0.041$, a positive-work deadband of $10^{-3}J$ gives $P(W>10^{-3}J)=0.5015>1/2$.

To test whether the branch-displacement diagnostic has energetic content, we resolve the negative-work probability and loss above several $K$ thresholds. Table~\ref{tab:tpm_tail_energetics} shows that the negative/positive tail enrichment increases steadily with threshold, while the same large-$K$ sectors carry a substantial fraction of the negative-work loss. The mean conditional loss also grows across broad displacement bins, from $-0.112J$ for $K<10$ to $-0.314J$ for $50\le K<60$, $-0.430J$ for $60\le K<80$, and $-0.831J$ for $K\ge100$. The lowest 32 hot-side branches contain $50.8\%$ of the Gibbs weight but account for $71.1\%$ of negative cycles and $61.4\%$ of the negative-work loss.

\begin{table*}[t]
\caption{\textbf{Threshold dependence of the negative-work displacement tail.}
Values refer to $L=233$, $T_h/J=0.65$, and $\nu/J=0.04$. The probability columns are conditional on work sign, and the last column is the fraction of the total negative-work loss carried by the indicated tail.}
\label{tab:tpm_tail_energetics}
\squeezetable
\begin{ruledtabular}
\begin{tabular}{lccc}
Threshold & $P(K\ge K_0\mid W<0)$ & $P(K\ge K_0\mid W>0)$ & Loss fraction \\
\hline
$K_0=20$ & 0.6610 & 0.5097 & 0.7854 \\
$K_0=40$ & 0.2455 & 0.1077 & 0.4278 \\
$K_0=50$ & 0.1328 & 0.0424 & 0.2798 \\
$K_0=60$ & 0.0673 & 0.0160 & 0.1706 \\
$K_0=80$ & 0.0154 & 0.00228 & 0.0519 \\
$K_0=100$ & 0.00234 & 0.000262 & 0.0103 \\
\end{tabular}
\end{ruledtabular}
\end{table*}

\section{Population contacts and stationary operation}
\label{app:contacts}

For a contact with bath \(b\in\{h,c\}\), the population map can be written
\begin{equation}
	\mathsf C_b=(1-\lambda_b)I+\lambda_b\bm p_{\beta_b}\bm 1^{\mathsf T},
	\qquad
	\lambda_b=1-e^{-\Gamma_b\tau_b}.
	\label{eq:contactmatrix}
\end{equation}
Here \(\tau_b\) is the contact duration and \(\Gamma_b\) the contact rate. The dimensionless contact coordinate is \(x_b=\Gamma_b\tau_b\), so the cycle time contains \(x_b/\Gamma_b\). For nonzero \(\lambda_b\), the Gibbs distribution is the unique fixed point of this semigroup; \(\lambda_b=1\) is a complete reset. Exactly zero contact on both baths is excluded because it does not select a unique stationary cycle.

With column probability vectors, one full cycle maps
\begin{equation}
	\bm p\longmapsto\mathsf M\bm p,\qquad
	\mathsf M=\mathsf C_hP_R\mathsf C_cP_F .
	\label{eq:stationarymap}
\end{equation}
The stationary population \(\bm p_*\) satisfies \(\mathsf M\bm p_*=\bm p_*\) and \(\bm 1^{\mathsf T}\bm p_*=1\). The reported critical contact strengths are the values at which the stationary mean work changes sign for the specified paths and this population semigroup.

For the coherence-sensitivity control, each contact instead acts on the density matrix in the corresponding endpoint energy basis as
\begin{equation}
\mathcal T_b[\rho]=(1-\lambda_b)\rho+\lambda_b\rho_{\beta_b}.
\label{eq:coherencecontact}
\end{equation}
The unitary strokes are propagated with the same Cayley--Crank--Nicolson amplitudes as the population calculation, and the affine cycle is iterated to its stationary density matrix. Equation~\eqref{eq:coherencecontact} retains the fraction $1-\lambda_b$ of pre-contact off-diagonal coherence and serves as a controlled coherence-retaining reference channel. At symmetric $\lambda=0.40$ on the extended path it gives $W/J=1.894\times10^{-3}$ compared with $-7.018\times10^{-4}$ from the population semigroup. On the crossing path the two models give negative work at $\lambda=0.83$ and positive work at $0.84$, bracketing their sign changes within the same interval. Propagator unitarity errors are below $10^{-14}$ in this control, and the stationary density matrices are iterated until the Frobenius-norm change is below $10^{-13}$. 

\section{Tilted cycle, long-time cumulants, and entropy}
\label{app:tilted}

Label a measured population trajectory through one cycle as \(i_h\to j_c\to k_c\to \ell_h\to m_h\). For a work counting field \(s_w\), the tilted work-stroke matrices are
\begin{align}
	[\mathsf F_{s_w}]_{ji}
	&=P_F(j|i)\exp[s_w(E_i^h-E_j^c)],\\
	[\mathsf R_{s_w}]_{\ell k}
	&=P_R(\ell|k)\exp[s_w(E_k^c-E_\ell^h)].
\end{align}
For heat counting fields \(s_c\) and \(s_h\), the tilted cold- and hot-contact matrices are
\begin{align}
	[\mathsf C_{s_c}]_{kj}
	&=[\mathsf C_c]_{kj}\exp[s_c(E_k^c-E_j^c)],\\
	[\mathsf H_{s_h}]_{m\ell}
	&=[\mathsf C_h]_{m\ell}\exp[s_h(E_m^h-E_\ell^h)].
\end{align}
These fields count the energy changes of the realised contact transitions; in particular, $s_c$ counts the physical cold-contact heat for $j_c\to k_c$. Using the column-vector convention of Eq.~\eqref{eq:contactmatrix}, the joint tilted one-cycle operator is
\begin{equation}
	\mathbb T_{\bm s}=\mathsf H_{s_h}\mathsf R_{s_w}
	\mathsf C_{s_c}\mathsf F_{s_w},\qquad
	\bm s=(s_w,s_h,s_c).
	\label{eq:tiltedcycle}
\end{equation}
The spectral radius of \(\mathbb T_{\bm s}\) gives the long-time scaled cumulant-generating function,
\begin{align}
	\psi(\bm s)&=\lim_{N\to\infty}\frac{1}{N}
	\ln\left\langle e^{s_w X_N^{(w)}+s_h X_N^{(h)}+s_c X_N^{(c)}}\right\rangle\nonumber\\
	&=\ln\rho_{\max}(\mathbb T_{\bm s}),
	\label{eq:jointscgf}
\end{align}
where \(X_N^{(w)}\), \(X_N^{(h)}\), and \(X_N^{(c)}\) are the accumulated work and heats over \(N\) cycles and \(\rho_{\max}\) is the largest-modulus eigenvalue. Derivatives at zero fields give the long-time cumulants per cycle. In particular, intercycle covariance terms shift the long-time variance per cycle away from the one-cycle variance whenever memory is appreciable.

For the TPM-resolved stationary population trajectory, the mean entropy production is
\begin{equation}
	\Sigma_{\mathrm{TPM}}=-\beta_hQ_h-\beta_cQ_c
	=\Sigma_{\mathrm{contact,pop}}+\Delta S_{\mathrm{work,deph}} .
	\label{eq:entropycorrection}
\end{equation}
Here \(\Sigma_{\mathrm{contact,pop}}\) is the contribution generated by population relaxation during the two contacts, and \(\Delta S_{\mathrm{work,deph}}\) records the additional entropy change associated with the energy-basis measurements and dephasing used to define the TPM work trajectory. An unmeasured coherent cycle has different entropy accounting because this projective dephasing step is absent. At the \(L=233\) crossing-path grid maximum-power point, the contact and measurement/dephasing contributions are approximately \(1.55\) and \(0.097\), giving \(\Sigma_{\mathrm{TPM}}\simeq1.65\). For the monitored population process, the Gibbs-renewal contacts obey local detailed balance and the retraced work strokes obey $P_R=P_F^{\mathsf T}$; the resulting entropy generating function satisfies the Gallavotti--Cohen symmetry over the evaluated counting-field range. This fluctuation symmetry refers specifically to the monitored population model; the coherence-retaining channel is treated separately. The joint work and heat statistics also satisfy the cycle first-law identity, with the difference between accumulated work and heat accounted for by the change in system energy at the initial and final cycle boundaries.
\section{Quasiperiodic parameter robustness}
\label{app:robustness}

The finite-rate sign and fluctuation patterns were examined under changes in the quasiperiodic phase, irrational frequency, rational approximants, and modulation balance. Across the eight phase values examined at the reference size, the principal stochastic pattern varies weakly. Replacing the inverse golden-mean frequency with one alternate irrational frequency preserves the qualitative mean/majority-sign separation at the selected path. Across these controls, the principal finite-rate sign separation varies weakly with phase offset and remains present for the alternate irrational frequency.

Rational Fibonacci approximants to the inverse golden-mean frequency change the fine minimum-gap sequence nonmonotonically, while the TPM observables approach their irrational-frequency values smoothly. The smooth convergence of the cycle statistics follows the Gibbs-weighted transition structure even as the extremal spacing varies nonmonotonically. Varying the irrational frequency can also change endpoint thermodynamics along the fixed path, and changing \(\theta\) changes the relative diagonal and off-diagonal modulation. Finite-rate accessibility consequently reflects the complete Hamiltonian path through the spectrum.

\section{Temperature, finite-window critical scaling, and return diagnostics}
\label{app:criticallimits}

The low-temperature analysis is performed by lowering the hot and cold reservoir temperatures together so that the complete-reset Gibbs states progressively concentrate on the lowest endpoint levels. In the joint ground-state limit the strict energy-ordered cycle returns the occupied branch to itself. For the two finite-rate ramps, the excess final energies above the corresponding energy-ordered branches are denoted \(\delta E_F\) and \(\delta E_R\). Direct evaluation verifies
\[
W_{\ad}-W_\nu=\delta E_F+\delta E_R
\]
to relative error approximately \(1.1\times10^{-6}\) for the evaluated low-temperature data.

Power-law fits in the accessible low-temperature rate window are treated as effective finite-window diagnostics. At $L=233$ and $T_c/J=0.006$, the full interval $0.006\leq\nu/J\leq0.030$ gives an exponent $0.48038$. Nested four-point windows give $0.40695$ on $0.006$--$0.014$, $0.46783$ on $0.008$--$0.020$, and $0.52877$ on $0.010$--$0.030$; the corresponding five-point fits are $0.44460$ and $0.50415$. Independent pure-state rate-window scans show the same systematic drift, while finite-size gap fits depend on whether the fixed irrational frequency or Fibonacci rational approximants are used. These trends identify a reproducible preasymptotic scaling regime and provide the quantitative basis for the finite-window Kibble--Zurek connection discussed in the main text.

Thermal broadening changes the effective scaling because more energy branches receive appreciable Gibbs weight. In the stored joint-temperature scan, the fitted work-deficit exponent is \(0.480\) at \(T_c/J=0.006\), where relatively few low-energy branches dominate, and rises to \(0.875\) at the baseline temperatures \(T_c/J=0.1\) and \(T_h/J=0.65\). At the higher baseline temperatures, the finite-rate response involves the full Gibbs-weighted transition and work statistics across many occupied branches.

For the rapid-driving comparison, the state-frozen reference is used as the sudden-quench endpoint map. The branch-resolved characteristic function develops a pronounced minimum on the crossing path. Across the four-size range, the continuous-ramp characteristic-function minimum saturates while the physical return remains finite and nearly size independent. The sudden reference carries the pronounced Loschmidt/characteristic-function structure, while the finite-rate ramps remain in a nonsingular crossover regime over the evaluated window.

\end{document}